\documentclass[aps,prb,twocolumn,showpacs]{revtex4-1}
\usepackage{graphicx,amssymb,amsfonts,amsmath}

\begin{document}

\title{Quench dynamics of one-dimensional bosons in a commensurate periodic potential:
A quantum kinetic equation approach}
\author{Marco Tavora and Aditi Mitra}
\affiliation{Department of Physics, New York University, 4 Washington
  Place, New York, New York 10003, USA}
\date{\today}

\pacs{05.70.Ln,37.10.Jk,71.10.Pm,03.75.Kk}

\begin{abstract}
Results are presented for the dynamics arising due to a sudden quench of a boson interaction parameter with the
simultaneous switching on of a commensurate periodic potential, the latter providing a source of non-linearity that
can cause inelastic scattering. A quantum kinetic equation is derived
perturbatively in the periodic potential and solved within the leading order gradient expansion.
A two-particle irreducible formalism is employed to construct the stress-momentum tensor and hence the conserved energy.
The dynamics is studied in detail in the phase where the boson spectrum remains gapless.
The periodic potential is found to give rise to multi-particle scattering processes that relaxes the boson
distribution function. At long times the system
is found to thermalize with a thermalization time that depends in a non-monotonic way on the amount of
energy injected into the system due to the quantum quench. This non-monotonic behavior arises due to the
competing effect of an increase of phase space for scattering on the one hand, and an enhancement of the orthogonality catastrophe
on the other hand as the quench amplitude is increased.
The approach to equilibrium is found to be purely exponential for large quench amplitudes,
and more complex for smaller quench amplitudes.
\end{abstract}

\maketitle

\section{Introduction}\label{inro}

How a many particle system prepared in an initial state which is far from equilibrium evolves in time,
and how it thermalizes if at all, is an important fundamental question which is also of experimental
relevance due to the realization of ideal, thermally isolated, quantum systems using ultra-cold atomic gases.~\cite{Bloch08} The issue
of thermalization becomes all the more important for one-dimensional (1D) and quasi 1D
systems where phase space for scattering is so highly restrictive that even generic systems with no special
conservation laws can show enormous bottlenecks to thermalization.~\cite{Weiss06,Gring12,Ronzheimer13} In fact in the past, classical
1D systems were also found to show a bottleneck to thermalization,~\cite{FPU} where a quantitative
explanation for this was found to be quite subtle, requiring an understanding of integrability, the Kolmogorov-Arnold-Moser (KAM) theorem and chaos.

One appealing way to understand thermalization or the lack of it in quantum systems is to borrow concepts
from classical systems.
In particular the lack of
thermalization in many 1D systems has been attributed to these systems being similar to
ideal integrable models.~\cite{Weiss06}
Integrable models have the property that they have many more conserved quantities besides energy and particle number,~\cite{Sutherland}
which strongly restricts their dynamics. However the exact effect
of integrability on dynamics is not understood in any quantitative way as the notion of integrability
cannot be generalized to quantum systems rigorously because classical and quantum systems
differ in the way the number of
degrees of freedom are counted.~\cite{Yuzbashyan11} This leads to confusion as to how to
collect all the relevant integrals of motion, and how many of these need to be kept in order to understand the
time-evolution and the long time behavior.~\cite{Caux11a} Recent studies have proposed keeping only the most ``local''
integrals of motion.~\cite{Essler13,Pozsgay13,Caux13} These selective integrals of motion have been used to
construct generalized Gibbs ensembles (GGEs)~\cite{Rigol07} with the aim to understand the long time steady-state behavior
after a quantum quench.
However understanding the temporal evolution from the initial state to the state described by the GGE in integrable models,
in particular those
described by interacting field theories is a challenging issue that has been addressed only recently for the time-evolution of some single-particle expectation values,~\cite{Fioretto10,Konik12,Mussardo13}
in contrast how two-point and higher order correlations evolve in time, and whether their behavior can be captured by a GGE is largely unexplored.

Numerical studies on finite size systems have also been extensively carried out to
understand the issue of thermalization.
The two main approaches used are exact diagonalization~\cite{Rigol09a,Rigol09b,Roux09,Biroli10,Santos10}
and integrability based methods.~\cite{Mossel12}
Some of these studies~\cite{Rigol09a,Rigol09b} indicate that
thermalization is consistent with the
eigenstate-thermalization-hypothesis (ETH),~\cite{Deutsch91,Srednicki94} and is related to the level statistics of
the Hamiltonian.~\cite{Santos10}
However it is a challenging task to generalize these results to systems in the thermodynamic
limit.

Perhaps one of the most powerful methods to study dynamics of many-particle systems in the thermodynamic limit is
the Schwinger-Keldysh or the quantum Boltzmann equation approach.~\cite{Kamenevrev,BergesRev} While the success of this approach relies
on the existence of a small parameter which allows one to truncate the self-energy to a finite order, it is still a non-perturbative
approach. These methods have been used to study nonequilibrium dynamics in a variety of interacting fermionic and
bosonic field theories, revealing thermalization for some cases,~\cite{Juchem04,AST61,Lindner08,Werner09,Gasenzer10} and
the appearance of intriguing non-thermal
quasi-stationary steady-states for other cases,~\cite{Berges04,Kehrein08,Kollar08,Berges08,Gasenzer10}
where some of these steady-states have been related to turbulence.~\cite{Gasenzer12}

A promising direction of research is to employ a quantum kinetic equation to study the dynamics of integrable
models, and to then study how integrability breaking terms affect the results. Recently such a study was undertaken for
the fermionic Hubbard model with nearest-neighbor~\cite{Spohn12} and next-nearest-neighbor~\cite{Spohn13}
interactions, where it was shown that the kinetic equation for the integrable case (nearest-neighbor interactions) allows
for an infinite number of athermal steady-states besides the thermal steady-state
corresponding to the Fermi-distribution function. Whether a system ends up in one of these athermal
steady-states or the thermal state depends on the initial condition.
Detailed studies also exist on how a 1D system weakly perturbed from an initial thermal equilibrium state relaxes to
equilibrium,~\cite{Andreev11,Glazman12,Pustilnik13} where the weak deviation from thermal equilibrium allows for a linearization of the Boltzmann equation.
However such a linearization procedure cannot be applied to quantum quenches since a quench
always generates a highly nonequilibrium distribution function.

The aim of the current paper is to study quench dynamics employing a quantum kinetic equation approach.
In particular we study a 1D system of interacting bosons in a commensurate periodic
potential. We envision a situation where the 1D bosons are initially in
the continuum (thus no periodic potential). Using Feshbach resonance techniques, at an initial time $t=0$, the interaction strength of the bosons is
suddenly quenched driving them out of equilibrium. A commensurate periodic potential is also switched on
at the same time as the interaction quench.
A periodic potential gives rise to Umpklapp or backscattering which can relax the distribution function of the bosons, however its effect
has never been studied before using a quantum kinetic equation approach. In this paper we plan to fill this gap. In particular
we employ a bosonization prescription to study the system, where the bosons in a commensurate periodic potential is described
by the sine-Gordon model.~\cite{Giamarchibook} Bosonization allows us to treat the effect of forward scattering between bosons exactly, while the
quantum kinetic equation is derived perturbatively in the Umpklapp or backscattering processes.

The ground state of bosons in a commensurate periodic
potential has two possible phases,~\cite{Giamarchibook} one is the superfluid phase where the periodic potential is an irrelevant
perturbation, and the low-energy theory of the bosons is described by the Luttinger liquid. The second phase is the Mott insulator
where the periodic potential is relevant and localizes the bosons. We will be interested in studying quenches within the
superfluid phase where the periodic potential is irrelevant in equilibrium. However out of equilibrium, irrelevant terms
cannot be dropped~\cite{Mitra11} and can eventually thermalize a system. We plan to study in detail how thermalization comes
about due to Umpklapp processes.
Of course, other irrelevant terms such as
band curvature are also present, however in this paper we retain only the irrelevant term corresponding to the commensurate
periodic potential. Keeping more than one irrelevant term is technically challenging, and beyond the scope of this paper.
Moreover, for certain quench parameters that we study, the cosine potential is the leading irrelevant operator, and therefore its effects
(such as inelastic scattering rate) will dominate
over other irrelevant operators.

Note that a periodic potential in a bosonization language is highly nonlinear in the bosonic
fields as it is given by $g \cos(2\phi)$, where $-\partial_x\phi/\pi$ is the
boson density. Since the system is a superfluid, $\phi$ is highly fluctuating, and the quench only increases these
fluctuations. Thus we are not allowed to Taylor expand the $\cos (2\phi)$ term. Neither are semi-classical approaches
such as the Truncated Wigner Approximation valid.~\cite{PolkovnikovTWA,Lancaster10b}
Instead we will derive a quantum kinetic equation perturbatively in
the strength of the periodic potential $g$. In doing so we will show that the periodic potential leads to an unusual kinetic equation that
even to ${\cal O}(g^2)$ leads to multi-particle scattering of bosons.
This is in marked contrast to studies on $\phi^4$
theory or fermionic models where the leading order terms in the kinetic equation capture two-particle~\cite{Spohn12,Spohn13} or three-particle scattering
processes.~\cite{Glazman12,Mazets08}

In this paper we present results for the dynamics of the 1D sine-Gordon model in its gapless phase
using a quantum kinetic equation approach. In contrast, in the gapped phase, the perturbative expansion in $g$
employed to derive the kinetic equation is not valid. The sine-Gordon model
is the continuum limit of the Bose-Hubbard model which is non-integrable. At the same time,
the sine-Gordon model is also the continuum limit of the nearest-neighbor XXZ spin-chain which is integrable. The continuum model is some
approximation of the lattice model and does not share all its features. While the sine-Gordon model in its gapped
phase is known to be integrable,~\cite{Rajaramanbook} whether it is integrable in the gapless phase is not a question straightforward
to address. The
reason is that the sine-Gordon model in the gapless phase has ultra-violate (UV) divergences that need to be regularized,
and the integrability or lack of it may very well depend on the short-distance physics and therefore the regularization scheme.
In this paper we follow a particular regularization scheme, and the final results, once expressed in terms of suitable dimensionless
units, does not depend on this regularization. However
since we always see thermalization, we think that
our model does not have any non-trivial conservation laws.
Thus our results are probably not relevant to the integrable XXZ chain, and are
more relevant for the non-integrable Bose-Hubbard model. However the precise connection between the results of this paper,
and the integrability or lack of it of the original lattice models needs further exploration.

This paper is organized as follows. In Section~\ref{model} we present the model that will be studied,
and briefly discuss the properties of the quench in the
absence of the periodic potential where the system reduces to the exactly solvable Luttinger model.
In Section~\ref{dyson} the Dyson
equation to leading non-trivial order in the periodic potential is derived, and simplified using a leading order gradient expansion.
In Section~\ref{EC} a two-particle
irreducible (2PI) formalism is used to show that the Dyson equation can equivalently be derived from a variational
approach applied to the 2PI action. We show that this approach also naturally leads to the derivation of a stress-momentum
tensor, and from that the conserved energy.
In Section~\ref{results} we present our results for the
time-evolution of the boson distribution function to leading order in the gradient expansion.
Here we discuss time-scales for thermalization and present an analytic estimate
for this time-scale. Finally in Section~\ref{conclu}
we present our conclusions. Some of the details of the calculations are relegated to
Appendices~\ref{derivb},~\ref{zeromode} and~\ref{inelasticrate}.

\section{Model} \label{model}

The Hamiltonian for interacting bosons in a periodic potential after bosonization is given by,
\begin{eqnarray}
H&=& H_0 + V_{\rm sg}\label{HSG0}\\
H_0 &=& \frac{u}{2\pi}\int dx \biggl\{K\left[\pi \Pi(x)\right]^2 + \frac{1}{K}\left[\partial_x \phi(x)\right]^2
\biggr\}
\label{HSG1}\\
V_{\rm{sg}} &=& - \frac{gu}{\alpha^2}\int dx \cos\left(\gamma \phi(x)\right)\label{HSG2}
\end{eqnarray}
where $H_0$ is the quadratic part which describes the Luttinger liquid or
long lived sound modes that propagate with velocity $u$. The density of the sound modes is $\rho  =-\partial_x\phi/\pi$, whereas
$\Pi= \partial_x \theta/\pi$ is the variable canonically conjugate to $\phi$,
 $\left[\phi,\partial_y\theta(y)\right] = i \pi \delta(x-y)$.
$V_{\rm sg}$ represents the commensurate periodic potential whose most important effect is a source of backscattering which can
localize the density modes via the well known Berezenskii-Kosterlitz-Thouless (BKT)
transition.~\cite{Giamarchibook} Note that we will use the convention that for bosons
\begin{eqnarray}
\gamma=2
\end{eqnarray}
while $K=1$ corresponds to hard-core bosons, and $K\rightarrow \infty$ represents non-interacting bosons.
In this notation, the critical point separating the Mott-insulating
phase from the superfluid phase, in the limit $g\rightarrow 0$, is located at $K=2$. In a renormalization group
language, the cosine potential is a relevant perturbation for $K <2$, and irrelevant otherwise, provided
$g$ is sufficiently small. In this paper we will be concerned with quench dynamics within the superfluid phase.

It is convenient to represent the  fields $\phi,\theta$ in terms of bosonic creation and annihilation operators
($b_p,b_p^{\dagger}$),~\cite{Giamarchibook}
\begin{eqnarray}
&&\phi(x) =
-(N_{R}+N_{L})\frac{\pi x}{L}\nonumber \\
&&-\frac{i\pi}{L}\sum_{p\neq0}\left(\frac{L|p|}{2\pi}\right)^{1/2}
\frac{1}{p}
e^{-\alpha|p|/2-ipx}\left(b_{p}^{\dagger} + b_{-p}\right)\\
&&\theta(x) =
(N_{R}-N_{L})\frac{\pi x}{L} \nonumber \\
&&+
\frac{i\pi}{L}\sum_{p\neq0}
\left(\frac{L|p|}{2\pi}\right)^{1/2}
\frac{1}{|p|}e^{-\alpha|p|/2-ipx}\left(b_{p}^{\dagger} - b_{-p}\right).
\end{eqnarray}
where
\begin{eqnarray}
\Lambda = u \alpha^{-1}
\end{eqnarray}
is an ultra-violet cutoff. Thus,
\begin{eqnarray}
H_0 = \sum_{p\neq 0} u|p|\eta_p^{\dagger} \eta_p
\end{eqnarray}
where $\eta$ and $b$ are related by the canonical transformation,
\begin{eqnarray}
\eta_{p} & = & \cosh\beta_1 b_{p} + \sinh\beta_1 b_{-p}^{\dagger},\\
\eta_{-p}^{\dagger} & = & \cosh\beta_1 b_{-p}^{\dagger} + \sinh\beta_1 b_{p}.
\end{eqnarray}
and $e^{-2\beta_1} = K, u = v_F/K$.

\subsection{Interaction quench in the Luttinger Model: Properties of the quadratic theory} \label{quench}

In this sub-section, let us assume that $g=0$, so that we have an exactly solvable theory,
namely the Luttinger model which captures the characteristic power-law decays of the Luttinger
liquid.
In equilibrium, the Luttinger liquid is an approximate low-energy description of more complicated models in their gapless phases such as the XXZ spin-chain and the Bose-Hubbard model.
Out of equilibrium, when $g \neq 0$, there is no reason to expect that the effective low energy or long wavelength
theory is described by the Luttinger liquid~\cite{Mitra11} as irrelevant operators cause inelastic scattering.
The aim of this paper is to explore these effects.

Dynamics arising either due to a sudden quench or due to more complicated time-dependent protocols have been
studied extensively in the literature for the Luttinger model.~\cite{Iucci09,Cazalilla06,Lancaster10,Perfetto06,Dora11,Rentrop12}
Here we present some results that we
will need for the $g\neq 0$ case.
Let us suppose that the system at time $t<0$ is a Luttinger liquid with
interaction parameter $K_0$ and velocity $u_0$, and therefore described by the Hamiltonian
\begin{eqnarray}
&&H_i = \frac{u_0}{2\pi}\int dx
\biggl\{K_0\left[\pi \Pi(x)\right]^2 + \frac{1}{K_0}\left[\partial_x \phi(x)\right]^2
\biggr\} \nonumber \\
&&= \sum_{p\neq 0} u_0 |p| \eta_p^{\dagger} \eta_p
\label{Hidef}
\end{eqnarray}
Let us suppose that at $t=0$ there is an interaction quench from $K_0\rightarrow K$
so that the time evolution from $t >0$ is due to
\begin{eqnarray}
&&H_f = \frac{u}{2\pi}\int dx
\biggl\{K\left[\pi \Pi(x)\right]^2 + \frac{1}{K}\left[\partial_x \phi(x)\right]^2
\biggr\} \nonumber \\
&&= \sum_{p\neq 0}u|p|\gamma_p^{\dagger}\gamma_p
\label{Hfdef}
\end{eqnarray}
To simplify the algebra we make the assumption that the quench preserves Galilean invariance
$u = v_F/K, u_0=v_F/K_0$. In the language of the Luttinger model this is equivalent to
always having $g_2$ and $g_4$ processes of the same magnitude.~\cite{Giamarchibook}
Note that the following canonical transformation relate the different sets of bosons,
\begin{eqnarray}
\begin{pmatrix} b_p \\ b_{-p}^{\dagger}\end{pmatrix}
= \begin{pmatrix} \cosh\beta_1 & -\sinh\beta_1 \\-\sinh\beta_1 &\cosh\beta_1
\end{pmatrix}
\begin{pmatrix} \gamma_p \\ \gamma_{-p}^{\dagger}\end{pmatrix}\\
\begin{pmatrix} b_p \\ b_{-p}^{\dagger}\end{pmatrix}
= \begin{pmatrix} \cosh\beta_0 & -\sinh\beta_0 \\-\sinh\beta_0 &\cosh\beta_0
\end{pmatrix}
\begin{pmatrix} \eta_p \\ \eta_{-p}^{\dagger}\end{pmatrix}
\end{eqnarray}
where $e^{-2\beta_0}=K_0, e^{-2\beta_1} = K$.

Let us define the functions
\begin{eqnarray}
f(pt) &&= \cos(u|p| t)\cosh\beta_0 \nonumber \\
&&-i \sin(u|p| t)\cosh(2\beta_1-\beta_0)\\
g(pt) &&= \cos(u| p| t)\sinh\beta_0 \nonumber \\
&&+ i \sin(u|p| t)\sinh(2\beta_1-\beta_0)
\end{eqnarray}
These functions determine the time-evolution after the quench ($t > 0$),
\begin{eqnarray}
b_p^{\dagger}(t) + b_{-p}(t) &&=\left(f^*(pt)-g(pt)\right)\eta_p^{\dagger}(0) \nonumber \\
&&+
\left(f(pt)-g^*(pt)\right)\eta_{-p}(0)\\
b_p^{\dagger}(t) - b_{-p}(t) &&=\left(f^*(pt)+g(pt)\right)\eta^{\dagger}_p(0) \nonumber \\
&&-\left(f(pt)+g^*(pt)\right)\eta_{-p}(0)
\end{eqnarray}
Note that the dynamics couples only the $q,-q$ modes but there is no scattering between
modes of different $|q|$. This will change when the periodic potential is applied in a way
that will be discussed in detail later. Moreover the dynamics is always translationally
invariant in position, both in the absence and presence of the periodic potential as long as
we are in the weak-coupling (in $g$) regime where the perturbative treatment of this paper is
valid. Of course looking for the growth of spatial instabilities after a quench is also an
interesting direction of study,~\cite{Foster11} which we do not address here.

\begin{figure}
\centering
\includegraphics[totalheight=5cm]{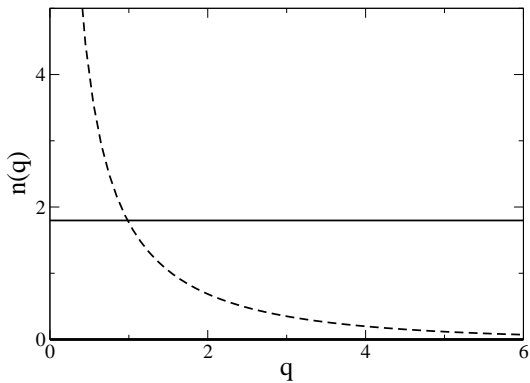}
\caption{The boson distribution function $n(q)$ generated by an interaction quench where $K_{\rm neq}=13.8$ and $K_{\rm eq}=3$ is
$n(q) = {1 \over 2}\left( {{{{K_{\rm neq}}} \over {{K_{\rm eq}}}} - 1} \right) \approx 1.797$ (continuous line).
This is compared with the boson distribution at equilibrium at a non-zero temperature
$1/\left[e^{u|q|/T_{\rm eq}}-1\right]$, where $T_{\rm eq}=2.23$ is the temperature associated with the
energy injected into the system due to the quench (dashed line). At zero temperature equilibrium,
$n(q) \equiv 0$.}
\label{ncomp}
\end{figure}

We will find it convenient to define the following two exponents,
\begin{eqnarray}
K_{\rm eq} = \frac{\gamma^2}{4}K \label{Keqdef}\\
K_{\rm neq} = \frac{\gamma^2}{8}K_0\left(1+ \frac{K^2}{K_0^2}\right) \label{Kneqdef}
\end{eqnarray}
In terms of these exponents the critical point separating the gapped and the gapless phases in
the limit $g \rightarrow 0$ is located at $K_{\rm eq}=2$. In the ground state of the final
Hamiltonian, correlators such as $\biggl \langle e^{i\gamma\phi(1)}e^{-i\gamma\phi(2)}\biggr\rangle$
decay as a power-law determined by $K_{\rm eq}$,
whereas at long times after the quench, the very same correlators decay as a power-law with the new
exponent $K_{\rm neq}$.~\cite{Cazalilla06,Mitra11} The origin of these new exponents is the underlying
integrability of the Luttinger model where the boson occupation number for each $q$ is conserved. A more quantitative
way to understand why after a quench the decays are still power-law is by noticing that the quench results in a
mode-occupation
\begin{eqnarray}
n(q)=\left\langle {\gamma _q^\dag {\gamma _q}} \right\rangle
= {{{{\left( {{K_0} - K} \right)}^2}} \over {4K{K_0}}} = \frac{1}{2}\left(\frac{K_{\rm neq}}{K_{\rm eq}}-1\right)\label{modeocc}
\end{eqnarray}
which is plotted in Fig.~\ref{ncomp}. This nonequilibrium distribution may be interpreted as one where there is
a momentum or mode dependent temperature. However since $n(q=0)$ is finite, it implies that the momentum dependent temperature
vanishes for long wavelengths. Since the power-law or lack of it is primarily determined by the mode-occupation or the effective
temperature of the long wavelength modes, the power law survives in the Luttinger model after a quench, though the
new exponent $K_{\rm neq}$ determines the decay.
As we show below, the fractional change in the decay exponent $\left[K_{\rm neq}-K_{\rm eq}\right]/K_{\rm eq}$ is a measure of how
far out of equilibrium the system is driven due to the quantum quench.

It is straightforward to see that the the mode-occupation in Eq.~(\ref{modeocc}) implies the following energy per unit length of the system,
\begin{eqnarray}
&&{E \over L} = {1 \over L}\sum\limits_{p \ne 0} e^{-\alpha |p|}
{u|p|\left\langle {\gamma _p^\dag {\gamma _p}} \right\rangle } \,\,\, = {u \over {4\pi {\alpha ^2}}}
{{{{\left( {{K_0} - K} \right)}^2}} \over {K{K_0}}}\nonumber \\
&&=\frac{u}{2\pi\alpha^2}\left(\frac{K_{\rm neq}}{K_{\rm eq}}-1\right)\label{En1}
\end{eqnarray}
Obviously the energy injected into the system depends on the underlying lattice cut-off $u/\alpha$. However once
length and energies are expressed in units of this cut-off, we have a cut-off independent result for the
energy injected into the system,
\begin{eqnarray}
\frac{\alpha E/u}{L/\alpha} =  \frac{1}{2\pi}\left(\frac{K_{\rm neq}}{K_{\rm eq}}-1\right)
\end{eqnarray}
Thus the energy injected into the system due to the quench is proportional to
the fractional change in the exponent governing the
power-law decay at long times, $\left[K_{\rm neq}-K_{\rm eq}\right]/K_{\rm eq}$. This is
an important energy scale which will determine the inelastic scattering rate
and therefore the thermalization time when the cosine potential is switched on.

In the presence of non-linearities and at long times, the system is expected to
eventually thermalize to the equilibrium (mode independent) temperature $T_{\rm eq}$. We now express
this temperature  in terms of the quench parameters
${{K_{{\rm{eq}}}}}$ and ${{K_{{\rm{neq}}}}}$.
Using $\left\langle {\gamma _p^\dag {\gamma _p}} \right\rangle  = {1 \over {{e^{u|p|/T_{\rm eq}}} - 1}}$, the energy
density at thermal equilibrium is,
\begin{eqnarray}
&&{E \over L} = {u \over {2\pi }}\left[ {{2 T_{\rm eq}\over {{u^2}}}\psi '
\left( {{\alpha  T_{\rm eq}\over {u}}} \right) - {2 \over {{\alpha ^2}}}} \right] 
\label{equilenergy}
\end{eqnarray}
where $\psi \left( z \right) = {{\Gamma '(z)} \over {\Gamma (z)}}$ and $\Gamma (z)$ is the gamma function. Since the system
is closed, energy conservation implies that
Eq.~(\ref{En1}) and Eq.~(\ref{equilenergy}) are equal to each other, so that $T_{\rm eq}$ is related as follows to the quench
parameters,
\begin{eqnarray}
&&{E \over L} = {u \over {2\pi }}\left[ {{2 T_{\rm eq}\over {{u^2}}}\psi '
\left( {{\alpha  T_{\rm eq}\over {u}}} \right) - {2 \over {{\alpha ^2}}}} \right] \nonumber \\
&& = \frac{u}{2\pi\alpha^2}\left(\frac{K_{\rm neq}}{K_{\rm eq}}-1\right)
\label{ee1}
\end{eqnarray}

Taking the limit of high temperatures, Eq.~(\ref{ee1}) becomes,
${E \over L} \approx
- {u \over {\pi {\alpha ^2}}} + {{{T_{{\rm{eq}}}}} \over {\pi \alpha }} $.
For high temperatures,
$\frac{\alpha}{u}T_{\rm eq}\gg 1$ (or large quenches $K_{\rm neq}\gg K_{\rm eq}$) energy conservation implies,
\begin{eqnarray}
\frac{\alpha}{u}{T_{{\rm{eq}}}} \approx {1\over {2}}{{{K_{{\rm{neq}}}}} \over {{K_{{\rm{eq}}}}}}
\label{TempLarge}
\end{eqnarray}
For small quenches $\left(K_{\rm neq}-K_{\rm eq}\right)/K_{\rm eq}\rightarrow 0$, and hence low temperatures,
the equilibrium temperature is given by,
\begin{eqnarray}
\frac{\alpha}{u}T_{\rm eq}\approx \frac{\sqrt{3}}{\pi}\sqrt{\frac{K_{\rm neq}}{K_{\rm eq}}-1}
\end{eqnarray}

Fig.~\ref{ncomp} shows the distribution function $n(q) = \langle \gamma_q^{\dagger}\gamma_{q}\rangle$
generated by an interaction quench and compares it with the
thermal distribution at temperature $T_{\rm eq}$.
For the Luttinger model, $n(q)$ is stable in time as there are no relaxation mechanisms. The aim
of this paper is to understand how the periodic potential relaxes the distribution function, and whether it ever reaches
the thermal distribution shown by the dashed line in Fig.~\ref{ncomp}.

\section{Periodic-potential: Derivation of the Dyson equation} \label{dyson}
We now turn to the case where a periodic potential is also present after the quench.
Thus the Hamiltonian at $t\leq 0$ is given by Eq.~(\ref{Hidef}), while that at $t>0$ is
given by Eqs.~(\ref{HSG0}),~(\ref{HSG1}),~(\ref{HSG2}). Thus at $t=0$ the boson interaction
parameter is quenched from $K_{0}\rightarrow K$, and a lattice of strength $g$ is also switched on.
Since we are interested in nonequilibrium dynamics, we will use the
Keldysh formalism.~\cite{Kamenevrev}
$\phi_{-/+}$ will denote fields that are (time/anti-time)-ordered on the Keldysh axes.
Accordingly we may define Green's functions $G_{ab}$, with $a,b=\pm$
where $G_{--}(12)= -i\langle T \phi_-(1)\phi_-(2)\rangle$ is the time-ordered Green's function,
$G_{++}(12)= -i \langle \tilde{T} \phi_+(1)\phi_+(2)\rangle$ is the anti-time ordered Green's function,
$G_{-+}(12)= -i\langle \phi_+(2)\phi_-(1)\rangle$ and $G_{+-}(12)= -i\langle \phi_+(1)\phi_-(2)\rangle$.

It is also convenient to
define quantum ($\phi_q$) and classical fields ($\phi_{cl}$),
\begin{eqnarray}
\phi_{\pm} = \frac{1}{\sqrt{2}}\left(\phi_{cl}\mp \phi_q\right)
\end{eqnarray}
with respect to which the basic retarded (R), advanced (A) and Keldysh (K) Green's functions are,
\begin{eqnarray}
G_R(xt,yt^{\prime})
&&= -i\theta(t-t^{\prime})\langle\left[\phi(xt),\phi(yt^{\prime})\right]\nonumber \\
&&=-i\langle \phi_{cl}(xt)
\phi_q(yt^{\prime})\rangle\\
G_A(xt,yt^{\prime})
&&= i\theta(t^{\prime}-t)\langle\left[\phi(xt),\phi(yt^{\prime})\right]\nonumber \\
&&=-i\langle \phi_{q}(xt)
\phi_{cl}(yt^{\prime})\rangle\rangle\\
G_K(xt,yt^{\prime}) &&= -i\langle \{\phi(xt),\phi(yt^{\prime})\}\rangle \nonumber \\
&&= -i \langle
\phi_{cl}(xt)
\phi_{cl}(yt^{\prime})\rangle
\end{eqnarray}
Note that $G_{R,A,K}$ are linear combinations of $G_{ab}$.
In what follows we will use lower case letters $g_{R,A,K},g_{ab}$ to denote the Green's functions for the
free theory, and upper-case for the Green's functions in the presence of the periodic potential.

The Green's function in the presence of a periodic potential is modified as follows
\begin{eqnarray}
&&i G_{cd}(1,2) = i g_{cd}(1,2)
- \frac{1}{2}\left(\frac{gu}{\alpha^2}\right)^2\sum_{a,b=\pm}\epsilon^{ab}\int d3 \int d4\nonumber \\
&&\langle \cos(\gamma\phi_a(3))\cos(\gamma\phi_b(4)) \phi_c(1)\phi_d(2)\rangle + \ldots
\end{eqnarray}
where $1,2,3,4$ denote both position and time indices, $\epsilon^{++}=\epsilon^{--}=1$ and
$\epsilon^{+-}=\epsilon^{-+}=-1$, while $\langle\rangle$ are expectation values evaluated in the
initial state which is the ground state of the free theory with interaction parameter $K_0$.
Note that the fact that the periodic potential was switched on suddenly at $t=0$ only appears in the
lower limit of the time integration. In general any switching protocol may be
employed.

\begin{figure}
\centering
\includegraphics[totalheight=5cm]{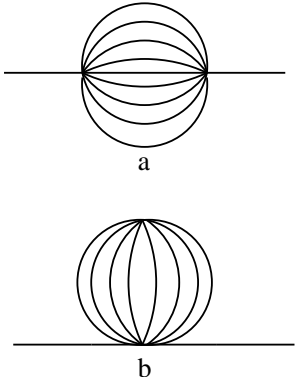}
\caption{The self-energies to leading order ${\cal O}(g^2)$ in the periodic potential. The solid lines are the boson Green's function $G$.}
\label{selfenergy}
\end{figure}

Performing the contractions, one finds,
\begin{eqnarray}
&&\langle \cos(\gamma\phi_a(3))\cos(\gamma\phi_b(4)) \phi_c(1)\phi_d(2)\rangle
\nonumber \\
&&=\langle \cos(\gamma\phi_a)\cos(\gamma\phi_b)\rangle\left(i g_{cd} +
\gamma^2\left[g_{ca}g_{ad}+g_{cb}g_{bd}\right. \right. \nonumber \\
&&\left. \left. -g_{ca}g_{bd}-g_{cb}g_{ad}\right]\right)
\end{eqnarray}
While the first disconnected term cancels when summing on indices $a,b$, the remaining terms
may be used to identify a self-energy $\Sigma$ to ${\cal O}(g^2)$ which is
shown diagrammatically in Fig.~\ref{selfenergy}, and whose formal expression is,
\begin{eqnarray}
\Sigma^R(1,2) = \Pi^R(1,2) -\delta(1-2)\int d3 \Pi^R(1,3)\\
\Sigma^K(1,2) = \Pi^K(1,2)
\end{eqnarray}
where,
\begin{eqnarray}
&&\Pi^R(x_1t_1,x_2t_2) = -i\gamma^2\left(\frac{gu}{\alpha^2}\right)^2\theta(t_1-t_2)
\nonumber \\
&&\left[\biggl\langle \cos(\gamma\phi_-(x_1t_1))\cos(\gamma\phi_-(x_2t_2))\biggr\rangle
\right. \nonumber \\
&&\left. - \biggl\langle \cos(\gamma\phi_+(x_1t_1))\cos(\gamma\phi_+(x_2t_2))\biggr\rangle
\right]\label{piR}\\
&&\Pi^K(x_1t_1,x_2t_2) = -i\gamma^2\left(\frac{gu}{\alpha^2}\right)^2
\nonumber \\
&&\left[\biggl\langle \cos(\gamma\phi_-(x_1t_1))\cos(\gamma\phi_-(x_2t_2))\biggr\rangle
\right. \nonumber \\
&&\left. + \biggl\langle \cos(\gamma\phi_+(x_1t_1))\cos(\gamma\phi_+(x_2t_2))\biggr\rangle
\right]\label{piK}
\end{eqnarray}
While the effect of this self-energy has been studied
in the past in order to understand equilibrium finite temperature
properties~\cite{Schulz86,Oshikawa02,Sirker11} of 1D systems, here we will be interested on
its effect on dynamics arising after a quench.

The Dyson equation obtained from the infinite summation of terms
$G = g + g\Sigma g + g \Sigma g \Sigma g + \ldots$ generalized to the time-dependent
problem, may now be written as,
\begin{eqnarray}
&&G_R(xt,yt^{\prime}) = g_R(xt,yt^{\prime}) + \int dx_1dt_1\int dx_2dt_2\nonumber \\
&& \times g_R(xt,x_1t_1)\Pi^R(x_1t_1,x_2t_2) \left[
 G_R(x_2t_2,yt^{\prime}) \right. \nonumber \\
&&\left. - G_R(x_1t_1,yt^{\prime}) \right]\nonumber \\
&&= g_R(xt,yt^{\prime}) + \int dx_1dt_1\int dx_2dt_2\nonumber \\
&& \times g_R(xt,x_1t_1)\Sigma^R(x_1t_1,x_2t_2)
 G_R(x_2t_2,yt^{\prime})
\label{DysonR}\\
&&G_K(xt,yt^{\prime}) = \int dx_1dt_1\int dx_2dt_2
G_R(xt,x_1t_1) \nonumber \\
&&\times \Pi^K(x_1t_1,x_2t_2)G_A(x_2t_2,yt^{\prime}) \label{DysonK}
\end{eqnarray}
where the retarded propagator for the Luttinger model is
\begin{eqnarray}
&&g_R^{-1}(1,2)= \nonumber \\
&&-\frac{1}{\pi K u}\left[\partial_{t_1}^2 - u^2\partial_{x_1}^2\right]\delta(t_1-t_2)\delta(x_1-x_2)
\end{eqnarray}

Since the Keldysh component $G_K$ contains information on the mode-occupation, its equation of motion constitutes a (quantum) kinetic equation.
An alternative way to write the Keldysh component is by expressing $G_K$ in the following way,~\cite{Kamenevbook}
\begin{eqnarray}
{G_K} \equiv {G_R} \circ F - F \circ {G_A}
\label{GKparametrized}
\end{eqnarray}
where the symbol $\circ$ implies convolution on $x$ and $t$ and also matrix multiplication.
Using the Dyson equation in matrix form $({{\hat g}^{ - 1}} - \hat \Sigma ) \circ \hat G = \hat 1$ we obtain the following kinetic equation in terms of the
auxiliary function $F$,
\begin{eqnarray}
F \circ g_A^{ - 1} - g_R^{ - 1} \circ F = {\Sigma _K} - {\Sigma _R} \circ F + F \circ {\Sigma _A} \label{kina}
\end{eqnarray}
Assuming spatial invariance which allows us to transform to momentum space,
and using that the left hand side of Eq.~(\ref{kina}) is $\frac{1}{\pi K u}\left[\partial_t^2 - \partial_{t^{\prime}}^2\right]F(q, t,t^{\prime})$,
and changing variables from $(t,t')$ to $(T,\tau ) = ((t + t')/2,t - t')$, Eq.~(\ref{kina}) becomes
\begin{eqnarray}
&{}& {\partial _\tau }{\partial _T}F(q,T,\tau ) = \left( {{{\pi Ku} \over 2}} \right)\left[ {{\Sigma ^K}(q,T,\tau )} \right.  \cr
&{}& \left. {\,\,\,\,\,\,\,\,\,\,\,\,\,\,\,\,\,\,\,\,\,\,\,\,\,\, - ({\Sigma ^R} \circ F)(q,T,\tau ) + (F \circ {\Sigma ^A})(q,T,\tau )} \right]
\end{eqnarray}
So far no approximations have been made other than in the precise diagrams that make up $\Sigma$.

We now define,
\begin{eqnarray}
\tilde F(q,T,\omega ) = \int\limits_{ - \infty }^\infty  {d\tau } \int\limits_{ - \infty }^\infty  {dr} {e^{ - iqr + i\omega \tau }}F(r,T,\tau )
\label{KinEq1}
\end{eqnarray}
and employ a gradient expansion to lowest order {\sl i.e.}, $(\widetilde {{\Sigma ^R} \circ F})(q,T,\omega ) \approx
{{\tilde \Sigma }^R}(q,T,\omega )\tilde F(q,T,\omega )$.~\cite{Rammer86}
Note that by dropping the derivatives in the gradient expansion we are assuming that the dynamics is valid for $t$ sufficiently
large since for small $t$, $F$ usually rapidly oscillates.~\cite{MuellerLindner06} This decoupling simplifies the kinetic equation to,
\begin{eqnarray}
&{}& {\partial _T}\tilde F(q,T,\omega ) = \left( {{{i\pi Ku} \over {2\omega }}} \right)\left[ {{{\tilde \Sigma }^K}(q,T,\omega )} \right.  \cr
&{}& \left. {\,\,\,\,\,\,\,\,\,\,\,\,\,\,\,\,\,\,\,\,\,\,\,\,\,\, - \left( {{{\tilde \Sigma }^R}(q,T,\omega ) - {{\tilde \Sigma }^A}(q,T,\omega )} \right)
\tilde F(q,T,\omega )} \right] \nonumber\\
\label{KinEq2}
\end{eqnarray}

Let us consider the analogous equation for the spectral density
\begin{eqnarray}
G_{\rho} = G_A-G_R\label{Grho}
\end{eqnarray}
The Dyson equations for
the retarded and advanced propagators are,
\begin{eqnarray}
&&\left(g_R^{-1}\circ G_R\right)(x x_0,y y_0) = 1 + \int dz \int_{y_0}^{x_0} dz_0\nonumber \\
&&\times \Sigma^R(xx_0,zz_0)G_R(zz_0,yy_0) \\
&&\left(g_A^{-1}\circ G_A\right)(x x_0,y y_0) = 1 + \int dz \int_{x_0}^{y_0} dz_0\nonumber \\
&&\times \Sigma^A(xx_0,zz_0)G_A(zz_0,yy_0)
\end{eqnarray}
where $x,y,z$ denote spatial coordinates and $x_0,y_0,z_0$ denote temporal coordinates.
Taking the difference of the above two equations and noting that terms such as $\theta(t_1-t_1^{\prime})
\int_{t_1^{\prime}}^{t_1} dt_2\Sigma^R(1,2) G^A(2,1^{\prime}) = 0$ we obtain,
\begin{eqnarray}
&&\left(g^{-1}\circ \right) G_{\rho}(x x_0,y y_0) = \int dz \int_{y_0}^{x_0} dz_0\nonumber \\
&&\times \left[\Sigma^R(xx_0,zz_0)-\Sigma^A(xx_0,zz_0)\right]G_{\rho}(zz_0,yy_0)
\end{eqnarray}
By applying the $g^{-1}$ operator on the second argument, and combining the result with the above equation, we obtain
\begin{eqnarray}
&&\left[g^{-1}\circ G_{\rho}- G_{\rho}\circ g^{-1}\right](xx_0,yy_0) = \int dz \int_{y_0}^{x_0} dz_0\nonumber \\
&&\times \left[\Sigma^R(xx_0,zz_0)-\Sigma^A(xx_0,zz_0)\right]G_{\rho}(zz_0,yy_0) \nonumber \\
&&- G_{\rho}(xx_0,zz_0)\left[\Sigma^R(zz_0,yy_0)-\Sigma^A(zz_0,yy_0)\right]
\end{eqnarray}
As before we now go into center of mass and relative coordinates, and assuming that the system is spatially homogeneous,
we find that to leading order in the gradient expansion,
\begin{eqnarray}
\omega {\partial _T}{{\tilde G}_\rho }(q,T,\omega ) = 0
\label{timeevolutionofsf}
\end{eqnarray}
In other words, the lowest order gradient expansion captures
the dynamics of the mode occupation and neglects the time evolution of the
spectral density, which is fixed by the state before the quench.
From the definition $~(\ref{GKparametrized})$
we note that $F$ comes multiplied by ${G_\rho}$ which is a sharply peaked function,
\begin{eqnarray}
{G_\rho }(q,\omega ) =  {{i{\pi ^2}K} \over {\left| q \right|}}\left[ {\delta (\omega  - u\left| q \right|) - \delta (\omega  + u\left| q \right|)} \right]
\label{onshell}
\end{eqnarray}
The kinetic equation can therefore be simplified by being computed only at the on-shell frequencies $\omega = \pm u q$
(below we simplify the notation by replacing $\tilde{\Sigma}, \tilde{F}$ by $\Sigma,F$),
\begin{eqnarray}
&&{\partial _T}F(q,T,\omega=uq) = \left( {{{i\pi K} \over {2q}}} \right)\left[ {\Sigma ^K}(q,T,\omega=u q) \right. \nonumber \\
&&\left. - \left({\Sigma ^R-\Sigma^A }\right)(q,T,\omega = u q)F(q,T, \omega = u q) \right]
\label{KE}
\end{eqnarray}
Thus the leading order gradient expansion is identical to the so called ``quasi-classical'' approximation which assumes that
the spectral density remains sharply peaked.

There is a hidden assumption in going into the mixed representation of $\omega,T$.
The time evolution of the system begins at $t=0$ but in order to be able to Fourier transform we need to
extend the range of the relative coordinate $\tau$ from $ - \infty$  to $\infty $ in $~(\ref{KinEq1})$.
Therefore by extending $t=0$ to $t=-\infty$ we are using a hybrid
description,~\cite{BergesBorsanyi06} since the initial conditions for the kinetic equation are defined at a finite time $t=0$.
This approximation
implies that the kinetic-equation cannot be used to describe dynamics at times too short after the quench at $t=0$,
but works best at slightly longer times where some of the memory effects have decayed away. A detailed comparison of the kinetic equation
with the full solution of the Dyson equation for $\phi^4$ theory and other models like the Yukawa model may be found in
Refs.~\onlinecite{MuellerLindner06,BergesBorsanyi06,BergesRev,Lindner08}.
Such a comparison for our model with the cosine potential
is beyond the scope of this paper.

The self-energies entering in the kinetic equation under the on-shell approximation for the spectral density are,
\begin{eqnarray}
&& {\Sigma ^K}(q,T,\omega ) =  - i{\left( {{{gu\gamma } \over {{\alpha ^2}}}} \right)^2}\int\limits_{ - \infty }^\infty  {d\tau }
\int\limits_{ - \infty }^\infty  {dr} \cos (qr)\cos (\omega \tau )  \nonumber \\
&& \,\,\,\,\,\,\,{e^{ - {I}(T,r,\tau )}}\cos \left[K_{\rm eq}\sum\limits_{\varepsilon  =  \pm 1} {{{\tan }^{ - 1}}
\left( {{{u\tau  + \varepsilon r} \over \alpha }} \right)} \right]  \label{SigK}\\
&& ({\Sigma ^R} - {\Sigma ^A})(q,T,\omega ) \cr
&&=  - i{\left( {{{gu\gamma } \over {{\alpha ^2}}}} \right)^2}\int\limits_{ - \infty }^\infty  {d\tau } \int\limits_{ - \infty }^\infty  {dr}
\sin (\omega \tau )\cos (qr)\,  \nonumber \\
&& \,\,\,\,\,\,{e^{ - {I}(T,r,\tau )}}\,\sin \left[ K_{\rm eq}\sum\limits_{\varepsilon  =  \pm 1} {{{\tan }^{ - 1}}
\left( {{{u\tau  + \varepsilon r} \over \alpha }} \right)} \right] \label{SigRA}
\end{eqnarray}
where,
\begin{eqnarray}
&{}& {I}(T,r,\tau ) = K_{\rm eq}\sum\limits_{\varepsilon  =  \pm 1} {\int\limits_0^\infty  {{{dq} \over q}} } {e^{ - \alpha q}}F(uq,T)  \cr
&{}& \,\,\,\,\,\,\,\,\,\,\,\,\,\,\,\,\,\,\,\,\,\,\,\,\,\,\,\,\,\,\,\,\,\,\,\,\,\,\,\,\,\,\,\,\,\,\left[ {1 - \cos
\left( {q\left( {u\tau  + \varepsilon r} \right)} \right)} \right] \label{Idef}
\end{eqnarray}
In what follows we will suppress the frequency label as it is understood that it is fixed at an on-shell value,
and use only the arguments $q$ and the time $T$ to label quantities.
The initial condition for the boson
distribution function follows from the discussion in Section~\ref{quench} and is given by
\begin{eqnarray}
F(q,T=0) = 1+ 2\left\langle {\gamma _q^\dag {\gamma _q}} \right\rangle(T=0)= \frac{K_{\rm neq}}{K_{\rm eq}}  \label{Fin}
\end{eqnarray}
In the absence of a quench, the distribution function is given by the zero temperature limit of $\coth\left(\frac{u|q|}{2T_{\rm eq}}\right)
\rightarrow 1$. Due to the interaction quench, a highly nonequilibrium boson distribution Eq.~(\ref{Fin}) is
generated. In the absence of the cosine potential this distribution function is infinitely long lived. However the cosine
potential allows for inelastic multi-particle scattering, which will relax it. We will study how this relaxation comes
about and whether the system thermalizes, and if so on what time-scales.

Eqs.~(\ref{KE}),~(\ref{SigK}),~(\ref{SigRA}),~(\ref{Idef}) together with the initial condition
in Eq.~(\ref{Fin}) define the problem we wish to solve. Note that the equilibrium
distribution function $F = \coth\frac{u|q|}{2T_{eq}}$
is a solution of the kinetic equation for any $T_{\rm eq}$, as it should be. Due to the self-energies being of the form
$\Sigma \sim \langle \cos(\gamma \phi)\cos(\gamma \phi)\rangle~\sim e^{-\gamma^2 \langle \phi\phi \rangle }$, where
$\langle \phi \phi\rangle$ is proportional to the boson distribution function,
the matrix elements determining the in-scattering and
out-scattering rates in the kinetic equation~(\ref{KE}) have an exponential dependence on the distribution function
through the function $e^{-I}\sim e^{-F}$.
This is in marked contrast to $\phi^4$ theory or other fermionic models
where the matrix elements entering the kinetic equation
have a polynomial dependence on the distribution function, for example for two-particle
scattering, the matrix elements
are proportional to $FF(1\pm F)(1\pm F)$. Hence we use the term multi-particle scattering to
describe the effect of the cosine potential.

It is instructive to study the form of the self-energy $\Sigma(r,0) \sim \langle \cos(\gamma \phi)(r)\cos(\gamma \phi)(0)\rangle$
at zero temperature equilibrium. Here $\Sigma\sim\frac{1}{|r|^{2K_{\rm eq}}}$. At long times after an interaction
quench, and within leading order perturbation theory where $F$ may be taken
to be the distribution function right after the quench (Eq.~(\ref{Fin})), $\Sigma \sim
\frac{1}{|r|^{2 K_{\rm neq}}}$. Since $K_{\rm neq} > K_{\rm eq}$, the matrix element for scattering is suppressed
by the interaction quench.

The above exponential dependence of the matrix elements on the distribution function is a consequence of the
orthogonality catastrophe arising due to the interaction quench. In particular the larger the quench amplitude
$K_{\rm neq}-K_{\rm eq}$, the poorer is the overlap between the initial wave-function
and the low-energy eigenstates of the final Hamiltonian, leading to
an exponential suppression of the matrix elements entering the kinetic equation.
Because of
our bosonization approach, this physics is captured naturally.
This also shows that a quench, however small in
magnitude, cannot lead to a suitable linearization procedure for the kinetic equation as is possible to do for small
deviations about thermal equilibrium.~\cite{Glazman12}

It is important to now summarize the conservation laws associated with Eq.~(\ref{KE}). The potential $\cos(\gamma \phi)$
does not conserve particle number, this is also the case for $\phi^4$ theory.~\cite{BergesRev}
However since our system is closed, energy is conserved. The total
energy has a kinetic contribution, and an interaction contribution arising from the cosine potential. In the
next section we carefully evaluate the conserved energy and prove that the kinetic equation~(\ref{KE})
obtained from the leading order gradient expansion is a conserving approximation. Readers not interested in the
derivation of the conserved energy may go straight to the results in Section~\ref{results}.

\section{2PI Formalism and Energy Conservation} \label{EC}
If we perform standard perturbative approximations for the self-energies$~(\ref{piR})$ and$~(\ref{piK})$,
such as replacing the dressed Green's function in the self-energy in Fig.~\ref{selfenergy}, by the bare one, we violate energy
conservation.~\cite{KadanoffBaym61} An efficient scheme to approach this problem and to
develop conserving approximations is via the two-particle irreducible (2PI)
effective action.~\cite{Ivanov98,Nishiyama,Gasenzer10} The 2PI effective action involves a diagrammatic series in
terms of closed loops where
dressed propagators are used instead of bare ones.~\cite{CornwallJackiwTomboulis}
By applying a variational principle on the 2PI functional, the exact Dyson
equations~(\ref{DysonR}) and~(\ref{DysonK}) are obtained.
The approximations are performed directly on the effective action by truncating the corresponding diagrammatic expansion.
By construction, the effective action (and consequently it's truncations) is invariant under global transformations of the Green's function
$G$, and using Noether's theorem we can find expressions for the conserved quantities
(in our specific case the energy).~\cite{AST61} The 2PI effective action $\Gamma(G)$ is given by,
\begin{eqnarray}
 &{}&\Gamma (G) \cr
&{}&= {i \over 2}\int {dxdt\int {dx'dt'} \ln ({G^{ - 1}})(xt,x't')\delta (x - x')\delta (t - t')}   \cr
 &{}&+ {i \over 2}\int {dxdt\int {dx'dt'} g^{ - 1}(xt,x't')G(x't',xt) + {\Gamma _2}(G)} \nonumber\\
\label{effectiveaction}
\end{eqnarray}
where $G(x,t;x',t')=\left\langle {{T_C}\varphi (x,t)\varphi (x',t')} \right\rangle$ is the Schwinger-Keldysh contour-ordered
Green's function and
$ig^{ - 1}(xt,x't') =-(\partial _t^2 - \partial _x^2)\delta (x - x')\delta (t - t')$ is the free propagator
(where we have set $u=1,\pi K=1$).
Since we are interested in finding the energy-momentum tensor ${T^{\mu \nu }}$ we apply a general (space and time dependent)
infinitesimal translation,
\begin{eqnarray}
x &\to& x + {\varepsilon_1}(x,t)  \cr
t &\to& t + {\varepsilon_0}(x,t)
\end{eqnarray}
which vanishes at the boundaries. To leading order in $\varepsilon(x,t)$ the corresponding transformation of $G(xt,x't')$ is
\begin{eqnarray}
&& \delta G(xt,x't')\nonumber \\
&&= {{\partial G(xt,x't')} \over {\partial x}}{\varepsilon _1}(x,t) + {{\partial G(xt,x't')} \over {\partial t}}{\varepsilon _0}(x,t)\nonumber \\
&&+ {{\partial G(xt,x't')} \over {\partial x'}}{\varepsilon _1}(x',t') + {{\partial G(xt,x't')} \over {\partial t'}}{\varepsilon _0}(x',t')\nonumber\\
&&= \left[\varepsilon_{\mu}(x,t)\partial_{\mu}+ \varepsilon_{\mu}(x',t')\partial_{\mu'}\right]G(xt,x't')
\label{eq:deltaG}
\end{eqnarray}
Above we used the notation ${x_\mu } = (x,t), x_{\mu'}=(x',t')$, and a sum over repeated indices is implied.

We will show below that we can write the variation of $\Gamma(G)$ as,
\begin{eqnarray}
\delta \Gamma (G) =  \int {dx\int {dt} \,{T^{\mu \nu }}} (x,t){{\tilde \partial }_\mu }{\varepsilon _\nu }(x,t)
\label{definitionofEMtensor}
\end{eqnarray}
The above equation defines the energy-momentum tensor $T^{\mu \nu }$.
The ${{\tilde \partial }_\mu }$ is defined as ${{\tilde \partial }_\mu } = ({{\tilde \partial }_x},{{\tilde \partial }_t}) \equiv
\left( { - {\partial _x},{\partial _t}} \right)$. After an integration by parts this becomes,
\begin{eqnarray}
 \delta \Gamma (G) = -\int {dxdt{{\tilde \partial }_\mu }{T^{\mu \nu }}} (x,t){\varepsilon _\nu }(x,t)
\label{ConservationofEMtensor}
\end{eqnarray}
Now, the Dyson equation is derived from the effective action $\Gamma(G)$ by requiring that
${{\delta \Gamma} \over {\delta G}}=0$. By taking the functional derivative of $\Gamma(G)$ we obtain,
\begin{eqnarray}
{{\delta \Gamma } \over {\delta G}} =  - {i \over 2}{G^{ - 1}}(xt,x't') + {i \over 2}g^{ - 1}(xt,x't')
+ {{\delta {\Gamma _2}} \over {\delta G}} \nonumber\\
\end{eqnarray}
Setting this to zero we obtain the Dyson equation,
\begin{eqnarray}
{G^{ - 1}}(x,t;x',t') - g^{ - 1}(x,t;x',t') + 2i{{\delta {\Gamma _2}} \over {\delta G}} = 0
\label{2PIaction}
\end{eqnarray}
provided that
\begin{eqnarray}
 \Sigma (x,t;x',t') = 2i{{\delta {\Gamma _2}} \over {\delta G}}
\end{eqnarray}
Thus we have,
\begin{eqnarray}
{G^{ - 1}}(xt,x't') = g^{ - 1}(xt,x't') - \Sigma (xt,x't')
\end{eqnarray}
Now we see that if $G$ satisfies the Dyson equation, Eqs.~(\ref{definitionofEMtensor}) and~(\ref{ConservationofEMtensor})
are zero. Moreover, since $({\varepsilon _0},{\varepsilon _1})$ is arbitrary, the energy-momentum tensor is
conserved
\begin{eqnarray}
{{\tilde \partial }_\mu }{T^{\mu \nu }} = 0
\end{eqnarray}
The above implies
\begin{eqnarray}
\partial_t T^{00} -\partial_xT^{x0}=0
\end{eqnarray}
Integrating over all space and noting that the second term above is a total derivative and therefore vanishes, leads
to the definition of the conserved energy $E=\int dx T^{00}$ such that,
\begin{eqnarray}
\frac{dE}{dt} = \frac{d}{dt}\int_{\rm all\,\, space} dx T^{00}(x,t) =0
\end{eqnarray}
For a spatially homogeneous system like the one we study, the energy density $T^{00}$ is position independent and is
also conserved. In this section we will
derive an explicit expression for $T^{00}$.

We can obtain the explicit expression for $T^{\mu \nu }$ by varying~$(\ref{effectiveaction})$.
We first define $\delta \Gamma (G) = \delta {\Gamma _a}(G) + \delta {\Gamma _b}(G) + \delta {\Gamma _2}(G)$ corresponding to varying the first, second and
third term in Eq.~(\ref{effectiveaction}).
We first vary ${\Gamma _a}(G)$
\begin{eqnarray}
\delta {\Gamma _a}(G) =  - {i \over 2}\delta {\rm{Tr}}\left( {\ln G} \right) =  - {i \over 2}{\rm{Tr}}{G^{ - 1}}\delta G
\label{eqn:functionalderivatives}
\end{eqnarray}
This can be written explicitly as,
\begin{eqnarray}
&{}&  - {i \over 2}\int {dxdt\int {dx'dt'} {G^{ - 1}}(xt,x't')}   \cr
&{}& \,\,\,\,\,\,\,\,\,\,\,\left[ {{\varepsilon _\mu }(x,t){\partial _\mu }G(x't',xt) + {\varepsilon _\mu }(x',t'){\partial _{\mu '}}G(x't',xt)} \right] \nonumber
\label{eqn:gamma2firstterm}
\end{eqnarray}
Introducing a $\delta$-function in the dummy variables $(x_2,t_2)$, we can rewrite the first term as (the second term is analogous),
\begin{eqnarray}
&{}&  - {i \over 2}\int {dxdt} \int {dx'dt'\int {d{x_2}d{t_2}} } \delta (x - {x_2})\delta (t - {t_2}){\varepsilon _\mu }({x_2},{t_2})  \cr
  &{}& \,\,\,\,\,\,\,\,\,\,\,\,\,{G^{ - 1}}(xt,x't'){\partial _{2\mu }}G(x't',{x_2}{t_2})  \cr
  &{}&  =  - {i \over 2}\int {dxdt} \int {d{x_2}d{t_2}} \delta (x - {x_2})\delta (t - {t_2}){\varepsilon _\mu }({x_2},{t_2})\,\,  \cr
  &{}& \,\,\,\,\,\,\,\,\,\,\,\,\,\,\,\,\,\,\,{\partial _{2\mu }}\int {dx'dt'} {G^{ - 1}}(xt,x't')G(x't',{x_2}{t_2})  \cr
  &{}&  =  - {i \over 2}\int {dxdt} \int {d{x_2}d{t_2}} \delta (x - {x_2})\delta (t - {t_2}){\varepsilon _\mu }({x_2},{t_2})\,\,  \cr
  &{}& \,\,\,\,\,\,\,\,\,\,\,\,\,\,\,\,\,\,\,\,\,\,\,\,\,\,\,\,\,\,\,\,\,{\partial _{2\mu }}\left[ {\delta (x - {x_2})\delta (t - {t_2})} \right]  \cr
  &{}&  =  - {i \over 2}\int {d{x_2}d{t_2}} {\varepsilon _\mu }({x_2},{t_2})\,{\partial _{2\mu }}\left[ {\delta (0)\delta (0)} \right]\, = 0
\end{eqnarray}
Treating the second term in a similar way we find $\delta \Gamma_a(G) =0$ and therefore does not contribute anything to the
stress tensor.

The expression for $\delta {\Gamma _b}(G)$ is,
\begin{eqnarray}
&{}& \delta {\Gamma _b}(G)  = {1 \over 2}\int {dx} dt\int {dx'dt'} \delta (x - x')\delta (t - t')(\partial _x^2 - \partial _t^2)  \cr
&{}& \left[ {{\varepsilon _1}(x,t){\partial _x}G(x't',xt)} \right. + {\varepsilon _1}(x',t'){\partial _{x'}}G(x't',xt)  \cr
&{}& \left. {\,\, + {\varepsilon _0}(x,t){\partial _t}G(x't',xt) + {\varepsilon _0}(x',t'){\partial _{t'}}G(x't',xt)} \right]
\end{eqnarray}
The above may be written in a short-hand form,
\begin{eqnarray}
&{}& \delta {\Gamma _b}(G)  \cr
&{}&  = {1 \over 2}\int {dx} dt\int {dx'dt'} \delta (x - x')\delta (t - t')  \cr
&{}& \,\,\,(\partial _x^2 - \partial _t^2)\left[ {{\varepsilon _\mu }(x,t){\partial _\mu }G(x't',xt) + {\varepsilon _\mu }(x',t')
{\partial _{\mu '}}G(x't',xt)} \right] \nonumber\\
\end{eqnarray}
If we now use the following expression
\begin{eqnarray}
 \partial _a^2({\varepsilon _\mu }{\partial _\mu }G)\mathop {{\rm{  }} = }\limits^{O(\varepsilon )} {\varepsilon _\mu }\partial _a^2({\partial _\mu }G) + 2({\partial _a}{\varepsilon _\mu }){\partial _a}({\partial _\mu }G)
\end{eqnarray}
where $\partial _a^2$ is $\partial _x^2$ or $\partial _t^2$ and perform a few manipulations we get,
\begin{eqnarray}
&{}& \delta {\Gamma _b}(G)  \cr
&{}&   = \int {dx} dt\int {dx'dt'} \delta (x - x')\delta (t - t')\,\,\,\,  \cr
&{}& \,\left[ {{\partial _t}{\varepsilon _\mu }(x,t)\,{\partial _{\mu '}}{\partial _t} - {\partial _x}{\varepsilon _\mu }(x,t)} \right.\,{\partial _{\mu '}}{\partial _x}\,  \cr
&{}& \left. {\,\,\,\,\,\, - {1 \over 2}{\partial _\mu }{\varepsilon _\mu }(x,t)\left( {\partial _{tt'}^2 - \partial _{xx'}^2} \right)} \right]G(x't',xt)\,
\end{eqnarray}
The above may be rewritten as (defining ${{\tilde \delta }_{\mu \nu }} = \left( { - {\delta _{xx}},{\delta _{tt}}} \right)$)
\begin{eqnarray}
&&\delta {\Gamma _b}(G) = \int {dx} dt\int {dx'dt'} \delta (x - x')\delta (t - t')\nonumber \\
&&\left[\left(\tilde{\partial}_{\nu}\epsilon_{\mu}\right)\partial_{\mu'}\partial_{\nu}-\frac{1}{2}\left(\tilde{\partial}_{\nu}\epsilon_{\mu}\right)\tilde{\delta}_{\mu\nu}
\left(\partial^2_{tt'}-\partial^2_{xx'}\right)\right]G(x't',xt)\nonumber\\
\end{eqnarray}
which has the required form~(\ref{definitionofEMtensor}).  We thus find
the following contribution to the energy-momentum tensor
\begin{eqnarray}
&{}& T_{\rm kin}^{\mu \nu }(x,t) = \int {dx'dt'\delta (x - x')\delta (t - t')}   \cr
&{}& \,\,\,\,\,\,\,\times \left[ {{{{\partial ^2}G(x',t';x,t)} \over {\partial {x_\mu }\partial {{x'}_\nu }}} - {1 \over 2}{{\tilde \delta }_{\mu \nu }}{{{\partial ^2}G(x',t';x,t)} \over {\partial {x_\lambda }\tilde{\partial} {{x'}_\lambda }}}} \right] \label{Tkin}
\end{eqnarray}
The label ${\rm kin}$ is to indicate that this is the contribution to the tensor for free bosons. In the presence of
interactions, even though we will continue to use the label ${\rm kin}$, $T_{\rm kin}^{\mu\nu}$ contains contributions from the interactions as the
Green's function $G$ is affected by the interactions. However in the presence of interactions, an additional contribution to the tensor arises from
varying the last term in Eq.~(\ref{effectiveaction}), which we discuss shortly.

Since the kinetic equation is expressed in terms of $F$ or equivalently $G_K$,
it is necessary to express the contour ordered Green's function $G$ in terms of $G_K$ and $G_\rho$.
Defining $\theta_C(t-t')$ to be the step function along the Keldysh contour, the decomposition identity is given by,~\cite{BergesRev}
\begin{eqnarray}
&{}& iG(x,t;x',t') = i\biggl\langle\phi (x,t)\phi (x',t')\biggr \rangle{\theta _C}(t - t') \cr
&{}& + i
\biggl\langle \phi (x',t')\phi (x,t)\biggr\rangle{\theta _C}(t' - t)  \cr
&{}&  = {1 \over 2}\overbrace { {i\biggl\langle\left\{  {\phi (x,t),\phi (x',t')} \right\}\biggr\rangle}}^{ - {G_K}(x,t;x',t')} +\cr
&{}&\,\,\,\,\,\,\,\,\,\,\,\,\,\,\,\,\,\,\,\, {1 \over 2}\overbrace { {i\biggl\langle {\left[ {\phi (x,t),\phi (x',t')} \right]}
\biggr\rangle } }^{{G_\rho }(x,t;x',t')}\overbrace {{\rm{sg}}{{\rm{n}}_C}(t - t')}^{{\theta _C}(t - t') - {\theta _C}(t' - t)}  \cr
&{}&  =  - {1 \over 2}\left[ {{G_K}(x,t;x',t') - {G_\rho }(x,t;x',t'){\rm{sg}}{{\rm{n}}_C}(t - t')} \right]
\label{decomposition}
\end{eqnarray}
and for the self-energy since $G^{-1}=g^{-1}-\Sigma$,
\begin{eqnarray}
&{}& i\Sigma (x,t;x',t') \cr
&{}&={1 \over 2}\left[ {{\Sigma ^K}(x,t;x',t') - {\Sigma ^\rho }(x,t;x',t'){\rm{sg}}{{\rm{n}}_{\rm{C}}}(t - t')} \right] \nonumber\\
\label{decomposition2}
\end{eqnarray}
Then we can write $T^{\mu \nu }_{\rm kin}(x,t)$ in terms of $G_K$ only since
\begin{eqnarray}
G(x,t;x,t) =   {i \over 2}{G_K}(x,t;x,t).
\end{eqnarray}
Equation~(\ref{Tkin}) becomes
\begin{eqnarray}
&& T^{\mu \nu }_{\rm kin}(x,t) = \int {dx'dt'\delta (x - x')\delta (t - t')}   \nonumber \\
&& \,\,\,\left[ {{i \over 2}{{{\partial ^2}{G_K}(x',t';x,t)} \over {\partial {x_\mu }\partial {{x'}_\nu }}} - {i \over 4}{{\tilde \delta }_{\mu \nu }}{{{\partial ^2}{G_K}(x',t';x,t)} \over {\partial {x_\lambda }\tilde{\partial} {{x'}_\lambda }}}} \right]\,\,\,\,\,\,\,\,\,\,\,\,\,\,\,\,\,\,\,\,\,\,\,\,\,\,\,
\end{eqnarray}
Thus the kinetic part of the energy density $T^{00}_{\rm kin}(x,t)$ after restoring the Luttinger liquid parameters reads,
\begin{eqnarray}
&& T^{00}_{\rm kin}(x,t) = \frac{1}{\pi K u}\int {dx'dt'\delta (x - x')\delta (t - t')} \,\,\,  \nonumber \\
&& \,\,\,\,\,\,\left[ {{i \over 4}{{{\partial ^2}{G_K}(x',t';x,t)} \over {\partial t\partial t'}} +
{i \over 4}u^2{{{\partial ^2}{G_K}(x',t';x,t)} \over {\partial x\partial x'}}} \right]
\end{eqnarray}

In consonance with the kinetic equation~(\ref{KE}) we perform a gradient expansion to lowest order, which as we showed before
is equivalent to the on-shell or quasi-classical approximation. This along with the assumption that the system
is homogeneous gives (defining $T=(t+t')/2,\tau=t-t'$),
\begin{eqnarray}
&& T_{{\rm{kin}}}^{00}(T) = \nonumber \\
&&{i \over (4\pi K u)}\int\limits_{ - \infty }^\infty  {{{dq} \over {2\pi }}} \int {d\tau\delta (\tau)}\nonumber \\
&& \times \left( \frac{1}{4}\partial_T^2- \partial_{\tau}^2 + {u^2q^2} \right){G_K}(q,T,\tau)   \nonumber \\
&& \approx {i \over (4\pi K u)}\int\limits_{ - \infty }^\infty  {{{dq} \over {2\pi }}} \int\limits_{ - \infty }^\infty
{{{d\omega } \over {2\pi }}} \left( {{\omega ^2} + {u^2q^2}} \right){G_K}(q,T,\omega )
\end{eqnarray}
where a term $O(\partial _T^2)$ was dropped.
Using
\begin{eqnarray}
&&{G_K}(q,\omega, T) =  \nonumber \\
&&- {{i{\pi ^2}K} \over {\left| q \right|}}\left[ {\delta (\omega  - u\left| q \right|)
- \delta (\omega  + u\left| q \right|)} \right]F(q,\omega,T)\nonumber\\
\end{eqnarray}
we obtain (suppressing the $\omega$ label as it is understood that it is evaluated at $\pm u q$)
\begin{eqnarray}
T_{{\rm{kin}}}^{00}(T) = {u \over {4\pi }}\int\limits_{ - \infty }^\infty  {dq} {e^{ - \alpha \left| q \right|}}\left| q \right|F(q,T)
\label{tkin}
\end{eqnarray}
This is the expected expression for the kinetic energy. In the absence of the cosine potential, this quantity is exact, and
was evaluated in Section~\ref{model}.
\begin{figure}
\centering
\includegraphics[totalheight=5cm]{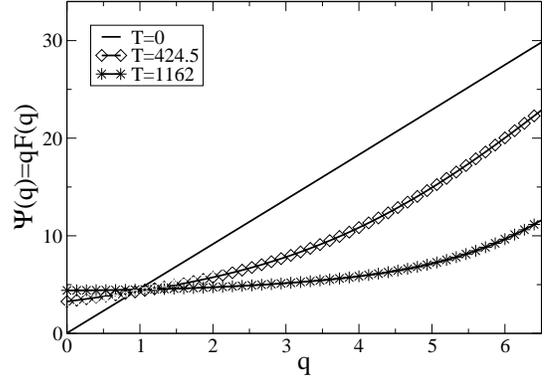}
\caption{The distribution function $\Psi (q) = qF(q)$ plotted for increasing times for a quench where $K_{\rm eq}=3,K_{\rm neq}=13.8$.
At the initial time,
the quench generates the distribution given by the straight line $\Psi (q) = q{{{K_{\rm neq}}} \over {{K_{\rm eq}}}}$.
The distribution converges to the equilibrium distribution given by $\Psi (q) = q\coth \left( {{{uq} \over {2{T_{\rm eq}}}}} \right)$.}
\label{FMcurves}
\end{figure}

Let us now construct the stress-momentum tensor from the third term in Eq.~(\ref{effectiveaction}),
and from that construct the interaction contribution
to the conserved energy.
The contribution from the last term $\Gamma_2$ can be written as,
\begin{eqnarray}
&{}& \delta {\Gamma _2}(G) = \int {dxdt\int {dx'dt'} } \overbrace {{{\delta {\Gamma _2}(G)} \over {\delta G(xt,x't')}}}^{(1/2i)\Sigma (x't',xt)}  \cr
&{}& \,\,\,\,\,\,\times \left[
{{\varepsilon _\mu }(x,t){\partial _\mu }G(xt;x't') + {\varepsilon _\mu }(x',t'){\partial _{\mu '}}
G(xt,x't')} \right]  \cr
&{}& \,\,\,\,\,\,\,\,\,\,\,\,\,\,\,\, = {1 \over {2i}}\int {dxdt\int {dx'dt'} } \Sigma (x',t';x,t)  \cr
&{}& \,\,\,\,\, \times \left[ {{\varepsilon _\mu }(x,t){\partial _\mu }G(xt;x't') + {\varepsilon _\mu }(x',t')
{\partial _{\mu '}}G(xt,x't')} \right]
\label{contributiongamma2}
\end{eqnarray}
Changing variables in the second term we obtain an expression of the form~(\ref{ConservationofEMtensor}),
\begin{eqnarray}
&{}& \delta {\Gamma _2}(G) = {1 \over {2i}}\int {dxdt\int {dx'dt'} } \left[ {\Sigma (x't',xt)\,{\partial _\mu }G(xt;x't') + } \right.  \cr
&{}& \left. {\,\,\,\,\,\,\,\,\,\,\,\,\,\,\,\,\,\,\,\,\,\,\,\,\,\,\,\,\,\,\,\,\,\,\,\Sigma (xt;x't'){\partial _\mu }G(x't',xt)} \right]{\varepsilon _\mu }(x,t)
\end{eqnarray}
where we identify
\begin{eqnarray}
&{}& {{\tilde \partial }_\mu }T_\Sigma ^{\mu \nu }(x,t) = -{1 \over {2i}}\int {dx'} dt'\left[ {\Sigma (x't',xt)\,{\partial _\nu }G(xt,x't') + } \right.  \cr
&{}& \left. {\,\,\,\,\,\,\,\,\,\,\,\,\,\,\,\,\,\,\,\,\,\,\,\,\,\,\,\,\,\,\,\,\,\,\,\,\,\,
\Sigma (xt,x't'){\partial _\nu }G(x't',xt)} \right]
\label{divergenceTmunu}
\end{eqnarray}
It is convenient to write the above expression in terms of $G_\rho$ and $G_K$ (see Appendix~\ref{derivb} for intermediate steps) to obtain,
\begin{eqnarray}
&{}& {{\tilde \partial }_\mu }T_\Sigma ^{\mu \nu }(x,t) = {1 \over {2i}}\int {dx'\int\limits_0^t {dt'} }
\left[ -{{\Sigma ^\rho }(x't',xt){\partial _\nu }{G_K}(xt,x't')} \right.  \cr
&{}& \left. {\,\,\,\,\,\,\,\,\,\,\,\,\,\,\,\,\,\,\,\,\,\,\,\,\,\,\,\,\,\,\,\,\,\,\,\,\,\,\,\, + {\Sigma ^K}(x't',xt){\partial _\nu }{G_\rho }(xt,x't')} \right]
\label{dTmunuKandrhocomp}
\end{eqnarray}
Integrating over $x$, the left hand side of the above equation is
\begin{eqnarray}
\!\!\!\!\!\int {dx} {{\tilde \partial }_\mu }T_\Sigma ^{\mu 0}(x,t) =\!\!\!\! \int {dx} {\partial _t}T_\Sigma ^{00}(x,t) -\!\!\!
\overbrace {\int {dx} {\partial _x}T_\Sigma ^{00}(x,t)}^{{\rm{surface\, terms\,}} \to {\rm{0}}}
\end{eqnarray}
Assuming spatial invariance, the expression for the interaction contribution to the conserved energy density $T_\Sigma ^{00}(t)$ is found to be,
\begin{eqnarray}
&& T_\Sigma ^{00}(t) =    {1 \over {2i}}\int\limits_0^t {dt''} \int\limits_0^{t''} {dt'} \int_{-\infty}^{\infty} {{{dq} \over {2\pi }}}{e^{ - \alpha \left| q \right|}}\nonumber \\
&&\left[ {{\Sigma ^K}(q,t',t''){\partial _{t''}}{G_\rho }(q,t'',t')} \right.  \nonumber \\
&&\left. - {\Sigma ^\rho }(q,t',t''){\partial _{t''}}{G_K}(q,t'',t') \right]
\end{eqnarray}

The next step is to perform a lowest order gradient expansion as we did before for the kinetic equation and for $T^{\mu\nu}_{\rm kin}$.
It is more convenient to work with ${dT_\Sigma ^{00}(t)/dt}$ and then integrate over $t$. The expression for ${dT_\Sigma ^{00}(t)/dt}$ is,
\begin{eqnarray}
&&{{dT_\Sigma ^{00}(t)} \over {dt}} = {1 \over {2i}}\int\limits_0^t {dt'} \int_{-\infty}^{\infty} {{{dq} \over {2\pi }}}{e^{ - \alpha \left| q \right|}} \nonumber \\
&&\left[ {{\Sigma ^K}(q,t',t){\partial _t}{G_\rho }(q,t,t')} \right.  \nonumber \\
&& \left. - {\Sigma ^\rho }(q,t',t){\partial _t}{G_K}(q,t,t') \right]
\end{eqnarray}
Rewriting this expression using Wigner coordinates $(T,\tau ) = \left( {{{t + t'} \over 2},t - t'} \right)$ one obtains,
\begin{eqnarray}
&& {{dT_\Sigma ^{00}(t)} \over {dt}} = {1 \over {2i}}\int_{-\infty}^{\infty}
{{{dq} \over {2\pi }}} \int\limits_0^t {dt'} {e^{ - \alpha \left| q \right|}}\nonumber \\
&&\left[{\Sigma ^K}\left( {q,{{(t + t')} \over 2},t'-t} \right) {\partial _t}{G_\rho }\left( {q,{{(t + t')} \over 2},t-t'} \right) \right. \nonumber \\
&&\left. - \left(\rho  \leftrightarrow K \right) \right],
\end{eqnarray}
The next step is to Fourier transform with respect to the relative coordinate $\tau=t-t'$.
As mentioned before, in order to go to $\omega$-space we must extend the interval of integration
of the relative coordinate
$\tau$ from $(0,t)$ to $(-\infty,\infty)$. Noticing that $\int\limits_0^\infty  {d\tau {e^{i(\omega  - \omega ')\tau }}}  \rightarrow \pi \delta (\omega  - \omega ')$,
using ${\partial _t}{G_\rho }(q,t,t') =  - i\int {{{d\omega } \over {2\pi }}{e^{ - i\omega (t - t')}}\omega {\tilde G_\rho }(q,T,\omega)\,} $
and finally integrating over $T$ we obtain the expression for $T_\Sigma ^{00}(T)$,
\begin{eqnarray}
&{}& T_\Sigma ^{00}(T) = -{1 \over 4}\int\limits_0^T {dT'} \int_{-\infty}^{\infty} {{{dq} \over {2\pi }}} {e^{ - \alpha \left| q \right|}}\int {{{d\omega }
\over {2\pi }}\omega } \left[ {{{\tilde \Sigma }^K}(q,T',\omega )} \right.  \cr
&{}& \,\,\,\,\,\,\,\,\,\,\,\,\,\,\,\,\,\,\,\,\,\,\,\,\,\,\,\,\,\,\,\,\,\,\,\,\,\,\,\,\,\,\,\,\,\,\,\,\,\,\,\,\, \times \left. {{{\tilde G}^\rho }(q,\omega )
- \left( {\rho  \leftrightarrow K} \right)} \right]
\end{eqnarray}
The final step is to use the on-shell approximation which as shown before follows from the leading order gradient expansion,
\begin{eqnarray}
&&  T_\Sigma ^{00}(T) =  - {1 \over 4}\int\limits_0^T {dT'} \int_{-\infty}^{\infty} {{{dq} \over {2\pi }}} {e^{ - \alpha \left| q \right|}}\int {{{d\omega } \over {2\pi }}\omega } \nonumber \\
&&\left[ {{{\tilde \Sigma }^K}(q,T',\omega ){{\tilde G}_\rho }(q,\omega )} \right.  \cr
&& \left. + {{\tilde \Sigma }^\rho }(q,T',\omega ){{\tilde G}_\rho }(q,\omega )F(q,T',\omega )
\right]  \nonumber \\
&&  =  - {{i{\pi ^2}K} \over {16{\pi ^2}}}\int\limits_0^T {dT'} \int_{-\infty}^{\infty} {dq} {e^{ - \alpha \left| q \right|}}\int {d\omega {\omega  \over {\left| q \right|}}}
\left[ {{{\tilde \Sigma }^K}(q,T',\omega ) + } \right.  \nonumber \\
&& \left. {\,\,{{\tilde \Sigma }^\rho }(q,T',\omega )\,F(q,T',\omega )} \right]\left[
{\delta (\omega  - u\left| q \right|) - \delta (\omega  + u\left| q \right|)} \right]
\nonumber \\
&&=  - {{iKu} \over 8}\int\limits_0^T {dT'} \int_{-\infty}^{\infty} {dq} {e^{ - \alpha \left| q \right|}}\left[ {{\tilde \Sigma }^K}(q,T') \right. \nonumber \\
&& \left. + {{\tilde \Sigma }^\rho }(q,T')\,F(q,T')
\right]
\end{eqnarray}

In summary the conserved energy density is
\begin{eqnarray}
\frac{E}{L} = T^{00}_{\rm kin}(T) + T^{00}_{\Sigma}(T)\label{Econs}
\end{eqnarray}
with $T^{00}_{\rm kin }$ defined in Eq.~(\ref{tkin}) and $T^{00}_{\Sigma}$ given by,
\begin{eqnarray}
&&T_{\Sigma}^{00}(T) = - {{iKu} \over 8}\int\limits_0^T {dT'} \int_{-\infty}^{\infty} {dq} {e^{ - \alpha \left| q \right|}}\left[ {\Sigma }^K(q,T') \right. \nonumber \\
&& \left. + { \Sigma }^\rho(q,T')\,F(q,T')
\right] \label{tsig}
\end{eqnarray}
On differentiating Eq.~(\ref{Econs}) in time and using the kinetic equation~(\ref{KE}),
we see that the total energy is conserved by the kinetic equation since,
\begin{eqnarray}
&{}& \frac{1}{L}{{dE(T)} \over {dT}} = {{iuK} \over 4}\int\limits_0^\infty  {dq} {e^{ - \alpha q}}\left[ {{\Sigma ^K}(q,T)
+ {\Sigma ^\rho }(q,T)F(q,T)} \right]  \cr
&{}&  - {{iuK} \over 4}\int\limits_0^\infty  {dq} {e^{ - \alpha q}}\left[ {{\Sigma ^K}(q,T)\,\, + \,{\Sigma ^\rho }(q,T)\,F(q,T)\,} \right]  \cr
&{}&  = 0
\end{eqnarray}
Thus the kinetic equation derived by us is a conserving approximation, with the conserved energy given in Eq.~(\ref{Econs}).

\section{Results} \label{results}

In this section we numerically solve the kinetic equation~(\ref{KE}) with the in-scattering and out-scattering rates given
in Eqs.~(\ref{SigK}),~(\ref{SigRA}),~(\ref{Idef}), and the boundary condition that at $T=0$ the distribution function is
given by Eq.~(\ref{Fin}).
As discussed in the previous sections, the cosine potential does not conserve particle number, but the quench always
conserves the total energy. Using the 2PI formalism we found the expression for
the conserved energy to be Eq.~(\ref{Econs}). It is the sum of a kinetic part Eq.~(\ref{tkin}) and an interaction
part Eq.~(\ref{tsig}). We find it convenient to
label the energies this way even though the interactions affect the kinetic part of the energy as well because they affect the
single-particle Green's function. In case the system
thermalizes, we would like to understand what is the equilibrium temperature at which it should thermalize. This temperature has a kinetic
contribution as well as a correction from the cosine interaction.

We find that after the leading order gradient expansion there is another zero mode in the problem, namely
\begin{eqnarray}
&&\frac{dT^{00}_{\rm kin}(T)}{dT}=0; \frac{dT^{00}_{\Sigma}(T)}{dT}=0 \label{zm}
\end{eqnarray}
The proof of this is given in Appendix~\ref{zeromode}. The above implies that under the leading order gradient
expansion, the total kinetic energy (summing over all momentum modes) and the total potential energy are separately
conserved. This additional conservation law implies that since at the initial time $T_{\Sigma}^{00}(T=0)=0$, it remains zero
always. This implies that the equilibrium temperature at which the system thermalizes can be calculated in a rather straightforward
way using the analysis of Section~\ref{quench} for the
quadratic Luttinger model. In particular the equilibrium temperature $T_{\rm eq}$ even in the
presence of the cosine potential is given by Eq.~(\ref{ee1}) within the leading order gradient
expansion.

The appearance of this zero mode where the total kinetic energy of the bosons is conserved is not an accident.
Recall that even for $\phi^4$ theories or interacting fermionic models, a kinetic equation obtained from
a leading order gradient and quasi-particle approximation results in
two and three particle scattering processes where the
total kinetic energy of the particles is conserved. The interactions at most
give a Hartree correction to the kinetic energy. While our kinetic equation is more complicated
than that for two and three particle scattering, yet the leading order gradient expansion leads to the same result.

All our numerical computations are done for $g=0.1$ and $\gamma=2$. Moreover we set the velocity $u=1$, and all
energy and length scales are in units of the cut-off $\Lambda = u/\alpha$. It is also interesting to note that the
strength of $g$ only appears as an over-all multiplying factor in the kinetic equation~(\ref{KE}) so that
as $g$ decreases, the time-evolution slows proportionally as $g^2$.

\begin{figure}
\centering
\includegraphics[totalheight=5cm]{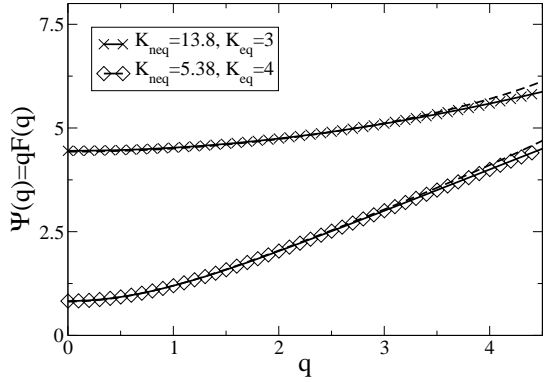}
\caption{At long times the distribution converges
to ${\Psi _{eq}}(q) = q\coth \left( {{{uq} \over {2{T_{\rm eq}}}}} \right)$ (dashed line). The quenches and the time for
which the distributions are shown correspond to
$K_{\rm eq}=3, K_{\rm neq}=13.8, T=1162$ and $K_{\rm eq}=4, K_{\rm neq}=5.38, T=1072$.
The final temperatures are respectively $T_{\rm eq}=2.22$ and $T_{\rm eq}=0.414$.}
\label{convergencetoequilibrium}
\end{figure}

It is convenient to define the function
\begin{eqnarray}
\Psi(q) = q F(q)
\end{eqnarray}
Since in equilibrium $\Psi(q) = q \coth\frac{uq}{2T_{\rm eq}}$, the $q=0$ intercept of $\Psi(q)$ may be used to define
an effective temperature,
\begin{eqnarray}
T_{\rm eff}= \frac{u}{2}\Psi(q=0)
\end{eqnarray}
Fig.~\ref{FMcurves} shows how the distribution function evolves in time. At $T=0$ the interaction quench
generates a nonequilibrium distribution function $\Psi(q,T=0) = q{{{K_{\rm neq}}} \over {{K_{\rm eq}}}}$.
The cosine potential leads to inelastic scattering that relaxes this
distribution function and generates an effective temperature $T_{\rm eff}$ which corresponds to
a non-zero intercept at $q=0$ in Fig.~\ref{FMcurves}. Eventually at long times this effective temperature evolves into
the true thermal temperature $T_{\rm eq}$. Fig.~\ref{convergencetoequilibrium} shows the distribution function at
sufficiently long times for two different quenches, and compares it with the equilibrium distribution function.
Not only the intercept at $q=0$, but also the
functional form of the distribution function agrees very well with the equilibrium form. There is some
deviation at large q, but we expect that this deviation will become smaller at longer times. To determine the
long distance behavior of various correlation functions, it is ultimately the distribution
function at small $q$ that is important.
Thus for all practical purposes,
Fig.~\ref{convergencetoequilibrium} is a fully thermalized distribution function.

\subsection{Toy model to recapture the dynamics for large quenches}

\begin{figure}
\centering
\includegraphics[totalheight=5cm]{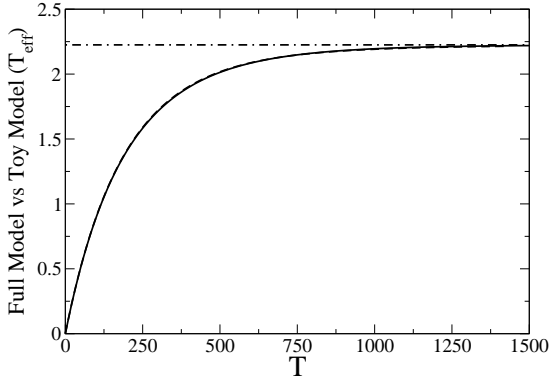}
\caption{Comparison between the full model (dashed line) and the toy model (solid line) for $K_{\rm eq}=3$ and $K_{\rm neq}=13.8$.
The differences are unobservable. The system thermalizes to $T_{\rm eq}$ (dot-dashed line). }
\label{TMvsFM}
\end{figure}
Simulating the full kinetic equation is numerically costly. Fig.~\ref{FMcurves} shows
that the distribution function stays mostly smooth as a function of $q$, and ultimately it is the
small $q$ behavior which is most important in determining the in-scattering and out-scattering rates. This
observation allows us to simulate the entire time-evolution using a toy model where we approximate the
distribution function by its value in the vicinity of $q=0$,
\begin{eqnarray}
\Psi(q>0,T)= 2 \frac{T_{\rm eff}(T)}{u} + B(T) q + A(T) q^2 + C(T) q^3\nonumber \\
\label{tm1}
\end{eqnarray}
where the coefficients $T_{\rm eff}, B,A,C$ are respectively the intercept, slope, and curvatures
of the distribution function at $q=0$. From
inversion symmetry $\Psi(q) = \Psi(-q)$. Substituting in Eq.~(\ref{KE}), the full
kinetic equation may be reduced to a set of coupled rate equations describing how
these four coefficients evolve in time, where the boundary conditions are
\begin{eqnarray}
&&A(T=0)=C(T=0)= T_{\rm eff}(T=0)=0; \nonumber \\
&&B(T=0) = \frac{K_{\rm neq}}{K_{\rm eq}}
\end{eqnarray}
The rate equations are found to be,
\begin{eqnarray}
&&\frac{dT_{\rm eff}(T)}{d T} = \frac{\pi K}{2}\left[\frac{u}{2}i \Sigma^K(q=0,T)\right. \nonumber \\
&&\left. -i\left(\Sigma^R-\Sigma^A\right)(q=0,T)T_{\rm eff}(T)\right]\\
&&\frac{d B(T)}{dT} = -\frac{\pi K}{2}i\left(\Sigma^R-\Sigma^A\right)(q=0,T)B(T)\\
&&\frac{d A(T)}{dT} = \frac{\pi K}{2}\left[-i\left(\Sigma^R-\Sigma^A\right)(q=0,T)A(T)\right. \nonumber \\
&&\left. + \frac{1}{2}\left(\frac{\partial^2i\Sigma^K}{\partial q^2}\middle|_{q=0}-\frac{2 T_{\rm eff}(T)}{u}
\frac{\partial^2 i\left(\Sigma^R-\Sigma^A\right)}{\partial q^2}\middle|_{q=0}\right)\right]\nonumber \\
\end{eqnarray}
where the coefficient $C$ is determined by imposing energy conservation and in particular $\frac{dT^{00}_{\Sigma}}{dT}=0$. The self-energies
and their momentum derivatives that appear above depend on the coefficients $T_{\rm eff}, A, B, C$
through their dependence on the distribution function entering in the exponent $I$ in Eqs.~(\ref{SigK})
and~(\ref{SigRA}). A comparison between the full solution of the
kinetic equation~(\ref{KE}) and the above toy model is shown in Fig.~\ref{TMvsFM}
where the time-evolution of the effective temperature after a quench is plotted. The two agree very well.
For smaller quench amplitudes however the deviation between the toy model and the full kinetic equation
becomes larger because for smaller quenches the dynamics, at least for short times, is affected by distribution
function at all $q$.

\begin{figure}
\centering
\includegraphics[totalheight=5.5cm]{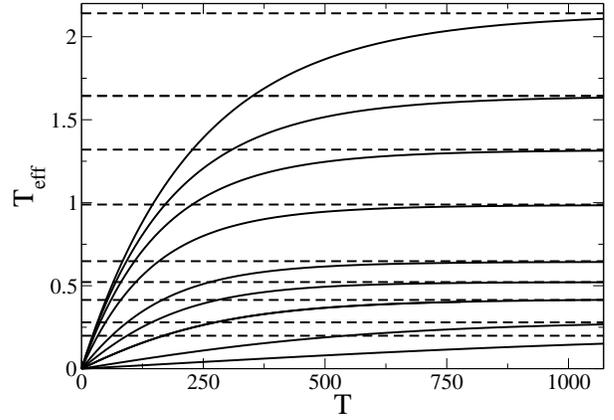}
\caption{Time-evolution of the effective temperature
for (bottom to top) $K_{\rm neq}=4.40,4.72,5.38,6.0,6.72,9.08,11.48,13.91$ and $17.72$ with $K_{\rm eq}=4$.
For all quenches the system thermalizes to the equilibrium temperate $T_{\rm eq}$ (dashed line).}
\label{allquenches}
\end{figure}

Fig.~\ref{allquenches} shows how the effective temperature evolves in time after a quench where $K_{\rm eq}=4$ while
$K_{\rm neq}$ is varied from $4.4$ to $17.72$. The larger quenches ($K_{\rm neq}\geq 6.0$) are obtained from the
toy model, while the smaller quenches are obtained from solving the full kinetic equation. For all cases,
the system is found to thermalize at the equilibrium temperature $T_{\rm eq}$ shown as a dashed line.

\subsection{Time-scales for thermalization}

An important question concerns the time-scales for thermalization. Assuming that the relaxation is purely exponential
$T_{\rm eff}(T)= T_{\rm eq}\left[1-e^{-\eta T}\right]$,
we define the thermalization time ($T_{\rm th}=1/\eta$) as the
time at which the effective temperature is related to the equilibrium temperature as follows,
\begin{eqnarray}
T_{\rm eff}(T_{\rm th}=1/\eta) = T_{\rm eq}\left[1-e^{-1}\right]\label{Thdef}
\end{eqnarray}
We find an interesting non-monotonic dependence of the thermalization time $1/\eta$ on the quench amplitude $K_{\rm neq}-K_{\rm eq}$. Naively
one expects the thermalization time to increase as the quench amplitude decreases, approaching infinity for zero quench amplitude
$K_{\rm eq}=K_{\rm neq}$. Note that a sudden switching on of $g$, keeping the interaction parameter fixed also generates nonequilibrium
dynamics, but this cannot be captured within the leading order gradient expansion.

\begin{figure}
\centering
\includegraphics[totalheight=5cm]{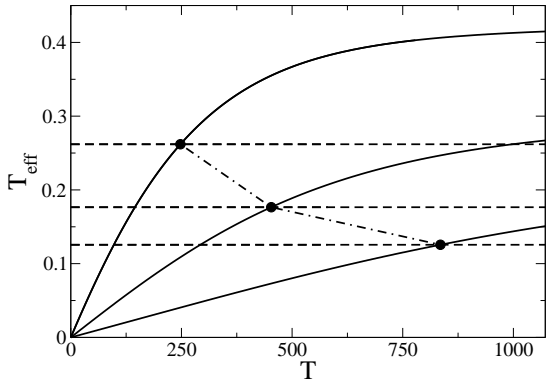}
\caption{For small quenches, the relaxation times (given by the abscissa corresponding to the three black dots)
decreases with increasing quench amplitude (dot-dashed line). The parameters are $K_{\rm eq}=4$ and $K_{\rm neq}=4.40, 4.72$ and $5.38$
(bottom to top).}
\label{relaxationtimessmallquenches}
\end{figure}

\begin{figure}
\centering
\includegraphics[totalheight=5cm]{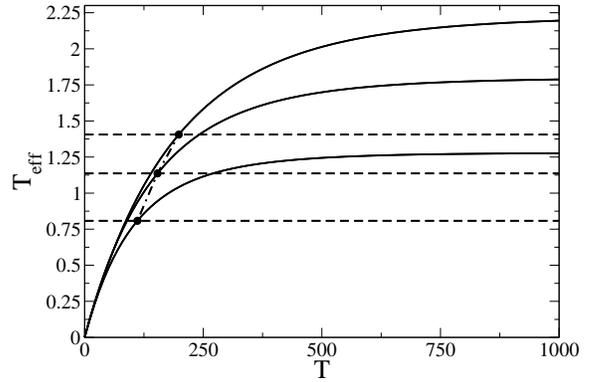}
\caption{For large quenches, the relaxation time (given by the abscissa corresponding to the three black dots)
increases with the quench amplitude (dot-dashed line). The parameters are $K_{\rm eq}=3$ and $K_{\rm neq}=8.39,11.3$ and $13.8$ (bottom to top).}
\label{relaxationtimeslargequenches}
\end{figure}

In contrast to the above expectation we find that the thermalization time increases with decreasing
quench amplitude (or the thermalization rate
increases with increasing quench amplitude) only for small quenches around equilibrium $K_{\rm neq}-K_{\rm eq}\lesssim K_{\rm eq}$.
For large quenches on the other hand
$K_{\rm neq}-K_{\rm eq}\gtrsim  K_{\rm eq}$, this behavior is reversed. Fig.~\ref{relaxationtimessmallquenches} shows
the expected increase in thermalization time for decreasing quench amplitudes for small quenches, while
Fig.~\ref{relaxationtimeslargequenches} shows how this behavior is reversed for larger quenches.

\begin{figure}
\centering
\includegraphics[totalheight=5.5cm]{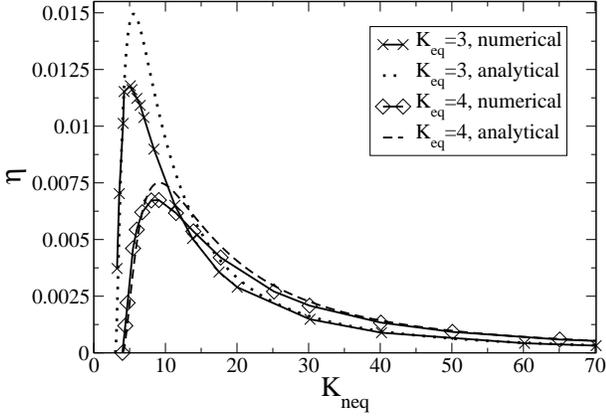}
\caption{The relaxation rate (inverse thermalization time) for quenches with different $K_{\rm neq}$ keeping $K_{\rm eq}$ fixed at
$K_{\rm eq}=3$ and $K_{\rm eq}=4$.
$\left[K_{\rm neq}-K_{\rm eq}\right]/K_{\rm eq}$ is proportional to the energy density injected due to the quench.
The numerical results
are compared with the analytic expression Eq.~(\ref{Ieta0}) obtained from perturbation theory.
}
\label{rates2}
\end{figure}

A summary of the relaxation rates (inversely related to the thermalization time) for quenches with $K_{\rm eq}=3$ and $K_{\rm eq}=4$
is shown in Fig.~\ref{rates2}, and clearly shows the non-monotonic behavior with a maximum relaxation rate for some
optimal $K_{\rm neq}$ for each $K_{\rm eq}$.
The non-monotonic behavior arises due to the dependence of the thermalization time on two competing effects. One
is the dependence of the scattering rate on the available phase-space for scattering that increases with increasing
quench amplitude and therefore favors an increase in the thermalization rate with increasing quench amplitude. The second
effect is that of the orthogonality catastrophe where a quench results in an initial state which has only a partial
overlap with the low-energy eigenstates of
the Hamiltonian after the quench, with the wavefunction overlap becoming poorer with increasing quench amplitude. This behavior
is reflected by the exponential dependence of the matrix elements entering the kinetic equation on the distribution function $F$.
Thus as the quench amplitude grows, $F$ increases, suppressing the scattering rates, and ultimately leading to
a non-monotonic dependence of the thermalization time on the quench amplitude.

It is interesting to ask how much of the numerical results may be recovered from a perturbative estimate for the
inelastic scattering rate. We use Eq.~(\ref{KE}) to
define an inelastic scattering rate,
\begin{eqnarray}
\eta_{0}=\!\! \lim_{\omega\rightarrow 0}\left(\frac{\pi K}{2}\right) \frac{i}{\omega}\left(\Sigma^R-\Sigma^A\right)(q,T,\omega=u q) \label{gamdef}
\end{eqnarray}
where $\Sigma^{R,A}$ depends on $F$ which is determined using the kinetic equation.
Such an inelastic scattering rate occurs in equilibrium and finite temperature as well
for weak deviations of the distribution function from thermal equilibrium, and
has been used to determine the finite-temperature lifetimes of the sound modes of the Luttinger liquid.~\cite{Sirker11}
We discuss the finite-temperature expression
in Appendix~\ref{inelasticrate}.

For our problem, we may make a crude estimate for the inelastic scattering rate by using perturbation theory, where
the distribution function $F$ entering in Eq.~(\ref{gamdef}) is taken to be the value right after the quench, $F =
\frac{K_{\rm neq}}{K_{\rm eq}}$.
This is equivalent to replacing the dressed Green's function in the self-energy by its bare value. While this is not a conserving
approximation, it still gives a result which shares many features with the results obtained
from the kinetic equation. As shown in Appendix~\ref{inelasticrate}, the perturbative estimate for $\eta_0$
in units of the cut-off is~\cite{Mitra11,Mitra12a}
\begin{eqnarray}
&&\eta_0 = \left(\frac{\pi K}{2}\right)g^2 \gamma^2\nonumber \\
&&\times \left[\frac{\pi}{2^{K_{\rm neq}-1}(K_{\rm neq}-1)
B\left(\frac{K_{\rm eq}+K_{\rm neq}}{2},\frac{K_{\rm neq}-K_{\rm eq}}{2}\right)}\right]\nonumber \\
&&\times \left[\frac{\pi}{2^{K_{\rm neq}-2}(K_{\rm neq}-2)
B\left(\frac{K_{\rm eq}+K_{\rm neq}-2}{2},\frac{K_{\rm neq}-K_{\rm eq}}{2}\right)}  \right. \nonumber \\
&&\left. - \frac{\pi}{2^{K_{\rm neq}-2}(K_{\rm neq}-2)
B\left(\frac{K_{\rm eq}+K_{\rm neq}}{2},\frac{K_{\rm neq}-K_{\rm eq}-2}{2}\right)}\right]\nonumber \\
\label{Ieta0}
\end{eqnarray}
Fig.~\ref{rates2} shows a comparison between the rate obtained from the kinetic equation, and
Eq.~(\ref{Ieta0}).
The agreement is impressive, especially for small and large quenches. The optimal value of $K_{\rm neq}$ at which
the rate is maximum is also in very good agreement.
For intermediate quench amplitudes,
a slight suppression of the actual relaxation rate compared with the perturbative estimate is found.

It is also interesting to ask how the thermalization time behaves as one approaches the critical point $K_{\rm eq}=2$. Fig.~\ref{rates2} shows
that the thermalization rates for quenches in the vicinity of $K_{\rm eq}=3$ is overall
greater than the thermalization rates in the vicinity of $K_{\rm eq}=4$.  Thus our results indicate that as one approaches the
critical point, the system relaxes faster. This observation has also been made in other theoretical and experimental
studies.~\cite{Cramer08,Trotzky12,Ronzheimer13}

It is instructive to see how the analytic expression in Eq.~(\ref{Ieta0}) behaves for small and large quench
amplitudes. For small quenches $K_{\rm neq}-K_{\rm eq}\ll 1$,
\begin{eqnarray}
&&\eta_0(K_{\rm neq}-K_{\rm eq}\ll 1) \simeq \nonumber \\
&&\frac{\pi^3K}{2} g^2 \gamma^2 \frac{2^{-2K_{\rm eq}}K_{\rm eq}}
{(K_{\rm eq}-2)(K_{\rm eq}-1)^2}\left(K_{\rm neq}-K_{\rm eq}\right)^2
\end{eqnarray}
Thus according to perturbation theory the thermalization rate diverges as one approaches
the critical point at $K_{\rm eq}=2$.
This divergence is unphysical and an indication that bare perturbation theory does not work
in the vicinity of the critical point, and a more self-consistent approach employing a kinetic equation
is necessary. In this paper we do not attempt to study the dynamics too close to the critical point using
the kinetic equation and leave it for future studies.
In the vicinity of the critical point, it may also
be necessary to use renormalization group
to improve on our perturbative expression for the self-energy because the cosine potential is a marginal
perturbation near the critical point.

Avoiding the above complications, away from the critical point, the thermalization rate obtained from perturbation theory is
found to behave as $g^2(K_{\rm neq}-K_{\rm eq})^2 \sim g^2 (K_0 -K)^4$ for small quenches, and is therefore proportional to
$g^2\times$ the square of the energy injected into the system due to the quench.
For large quenches $K_{\rm neq}-K_{\rm eq} \gg 1$, the thermalization rate is found to decrease with $K_{\rm neq}$ as
\begin{eqnarray}
\eta_0(K_{\rm neq}-K_{\rm eq}\gg 1) \simeq \pi g^2 \gamma^2\frac{K_{\rm eq}}{K_{\rm neq}^2}
\end{eqnarray}

\begin{figure}
\centering
\includegraphics[totalheight=5cm]{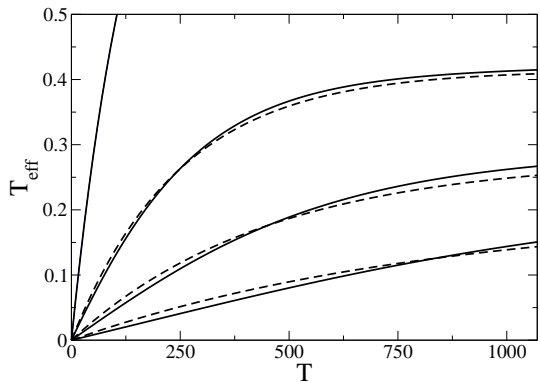}
\caption{Comparison between $T_{\rm{eff}}(T)$ obtained from the kinetic equation (solid line) with a relaxational ansazt
$T_{{\rm{eff}}}^{\exp }(T) = {T_{{\rm{eq}}}}\left( {1 - {e^{ - \eta T}}} \right)$ (dashed lines).
The parameters are $K_{\rm eq}=4$ and ${K_{\rm{neq}}} = 4.4,4.72,5.38$ and $9.08$ (bottom to top). As the quench amplitude decreases,
the deviation from a simple exponential relaxation becomes larger.
}
\label{comparisonwithexponentials}
\end{figure}

We now address the question of whether the approach to thermal equilibrium is truly exponential or not. Fig.~\ref{comparisonwithexponentials}
shows how the actual time-evolution of the effective temperature compares with a purely relaxational model
$T_{\rm eff}^{\rm exp}(T)= T_{\rm eq}\left[1-e^{-\eta T}\right]$. For large quenches, the differences between the two
are unobservable. However as the quench amplitude is reduced, the time-evolution deviates more and more from
a purely exponential relaxation. The thermalization time
is not precisely defined when the relaxation is not exponential, and so our definition for the thermalization time
in Eq.~(\ref{Thdef}) becomes somewhat ad hoc for small quenches.
Yet the close agreement with the analytic result shows that the definition
chosen is still a close estimate of the physical time scale.

\section{Conclusions and Outlook}\label{conclu}

We have presented a detailed study of quench dynamics of a 1D system of interacting bosons in a commensurate periodic
potential. A quantum kinetic equation was derived perturbatively in the strength of the periodic
potential and solved within a leading order gradient expansion. Our results are valid for quenches within the gapless (superfluid) phase.
The system is found to thermalize at long times. The
thermalization time is found to be non-monotonic in the quench amplitude or the amount of energy
injected into the system. This non-monotonic behavior
arises due to a competition between two effects. One is an increase in phase space for scattering as the system is driven further away from
equilibrium with increasing quench amplitude, this has the effect of relaxing the system faster as the quench
amplitude is increased. The second
is an enhancement of the orthogonality catastrophe with increasing quench amplitude arising out of a poorer overlap between
the initial wavefunction and the low-energy eigenstates of the final Hamiltonian as the quench amplitude is increased. This has the effect
of suppressing the matrix elements for scattering as the quench amplitude is increased,
leading to a longer thermalization time. The bosonization approach
captures the orthogonality catastrophe naturally. We also find an analytic expression for the
relaxation rate from perturbation theory which captures the above features remarkably well.

The result for the non-monotonic dependence of the thermalization time on the quench amplitude can also be understood as follows. Imagine that initially the bosons are in the ground state of a superfluid with interaction parameter $K_0$. Now consider increasing the quench amplitude to smaller and smaller values of $K$.
Here one would eventually be in
the Mott insulator phase where we expect thermalization to be poor or almost absent.
A similar behavior is expected when the quench amplitude is increased to larger and larger
values of $K$. Here since for $K=\infty$ the bosons
are effectively non-interacting, we again expect thermalization to be almost absent for large enough
$K$. This non-monotonic behavior of the thermalization time should be observable in experiments if a large tunability of quench
amplitudes were possible.

We find that in general the relaxation rates grow as one approaches the critical point. Previous numerical
and experimental studies which could approach the critical point both from the superfluid side as well as the Mott
insulating side made a similar observation, namely that the relaxation rates are maximal at the critical point and decrease
away from it in both directions.~\cite{Cramer08,Trotzky12,Ronzheimer13}

There are many open questions. Firstly we have employed a leading order gradient expansion which
works best when the relaxation rates are not too small so that the system looses memory of the initial condition fast.
Therefore this approach will not work when $K_{\rm neq}\rightarrow K_{\rm eq}$. In particular the leading order gradient expansion
gives no dynamics when only the periodic potential is quenched keeping the boson interaction parameter fixed. For this case a
full solution of the Dyson equation may be necessary, however this is a numerically challenging task.

The results of this paper are not valid very close to the critical point either. This is because our self-energies
were derived perturbatively in the periodic potential, whereas near the critical point, the periodic potential becomes
a marginal perturbation. For this
case an alternate approach presented in Ref.~\onlinecite{Mitra13a} may be more appropriate where a separation of time-scales
was identified. For times smaller than an inelastic scattering rate (but longer than microscopic time-scales),
real quantum processes dominate the dynamics and can be treated with a
perturbative renormalization group approach, while at longer times the quasi-classical approach
of this paper may be employed.

Finally the issue of integrability and the relation of this model to lattice models such as the Bose-Hubbard model
and the XXZ chain are open questions. Numerical studies on integrable and non-integrable fermionic models on the lattice show
that irrelevant operators do not affect the dynamics for numerically accessible time-scales.~\cite{Karrasch12}
Whether there is a fundamental difference between dynamics of lattice models and continuum field theories
like the one studied in this paper, or whether this is an issue of differing time-scales is a question
that needs to be explored.

Quench dynamics of integrable models described by interacting field theories, and the effect of integrability breaking
terms on the dynamics, is an important topic of research. Kinetic equations constructed for some 1D integrable models such as the nearest-neighbor fermionic Hubbard chain,~\cite{Spohn12} fermionic models with delta-function interactions in
real space,~\cite{Lunde07} the fermionic limit of the Lieb-Liniger model,~\cite{Lunde07} and the
Calogero-Sutherland model~\cite{Khodas07} have been shown to
know about the underlying integrability of the systems by either giving vanishing matrix elements for scattering,~\cite{Lunde07,Khodas07} or by causing scattering only in some special points, resulting in nonthermal steady-state solutions.~\cite{Spohn12} A more systematic
study involving other integrable models, and then understanding the effect of integrability breaking terms
is an important direction of research.

{\sl Acknowledgments}:
The authors thank I. Aleiner, J. Berges, F. Essler, S. Kehrein and M. Schir{\'o} for helpful discussions.
This work was partially supported by a grant from the Simons Foundation
and by the National Science Foundation under Grants No. PHY 11-25915 and No. DMR-1004589.


%

\appendix

\section{Derivation of Eq.~(\ref{dTmunuKandrhocomp})} \label{derivb}
Consider a Keldysh contour ($\int\limits_C$)
starting from an initial time $t_0$, going upto a maximum time $t_{\rm max}$ and then
returning to the initial time $t_0$. Then the following identities hold~\cite{BergesRev}
\begin{eqnarray}
&{}& \int\limits_C {dt'} {\rm{sg}}{{\rm{n}}_C}(t - t') = \int\limits_{{t_0}}^t {dt'}  - \int\limits_t^{{t_0}} {dt'}  = 2\int\limits_{{t_0}}^t {dt'}   \cr
&{}& \int\limits_C {dt'} {\rm{sg}}{{\rm{n}}_C}(t - t'){\rm{sg}}{{\rm{n}}_C}(t' - t'') =2{\rm{sg}}{{\rm{n}}_C}(t - t'')\int\limits_{t''}^t {dt'}
\nonumber\\
\label{ids}
\end{eqnarray}
The RHS of Eq.~(\ref{divergenceTmunu}) can be written in terms of $K$ and $\rho$ components by using the above identities.
The first term is
\begin{eqnarray}
&& {1 \over {2i}}\int {dx'} dt'\Sigma (x't',xt)\,{\partial _\mu }G(xt,x't')  =\nonumber \\
&&{1 \over {8i}}\int {dx'} dt'
\left[{\Sigma ^K}(x't',xt)
- {\Sigma ^\rho }(x't',xt){\rm{sg}}{{\rm{n}}_{\rm{C}}}(t' - t)\right] \nonumber \\
&& \times \left[ {{\partial _\mu }{G_K}(xt,x't') - {\partial _\mu }{G_\rho }(xt,x't'){\rm{sg}}{{\rm{n}}_C}(t - t')} \right]
\end{eqnarray}
where ${G_\rho }(xt,x't'){\partial _t}{\rm{sg}}{{\rm{n}}_C}(t - t') = 2{G_\rho }(xt,x't'){\delta_C}(t - t') = 0$ was used. Now using identities~(\ref{ids}) and
\begin{eqnarray}
&{}& \int\limits_C {dt'} {\Sigma ^K}(x't',xt){\partial _\mu }{G_K}(xt,x't')  \cr
&{}&  = \int\limits_{{t_0}}^{{t_{\max }}} {dt'} {\Sigma ^K}(x't',xt){\partial _\mu }{G_K}(xt,x't')  \cr
&{}&  + \int\limits_{{t_{\max }}}^{{t_0}} {dt'} {\Sigma ^K}(x't',xt){\partial _\mu }{G_K}(xt,x't')  \cr
&{}&  = 0
\end{eqnarray}
we end up with two terms only
\begin{eqnarray}
&{}& {1 \over {8i}}\int {dx'} dt'\left[ { - {\Sigma ^K}(x't',xt){\partial _\mu }{G_\rho }(xt,x't'){\rm{sg}}{{\rm{n}}_C}(t - t')} \right.  \cr
&{}& \left. { - {\Sigma ^\rho }(x't',xt){\partial _\mu }{G_K}(xt,x't'){\rm{sg}}{{\rm{n}}_{\rm{C}}}(t' - t)} \right]\,  \cr
&{}&  = -{1 \over {4i}}\left( {\int {dx'} \int\limits_0^t {dt'} {\Sigma ^K}(x't',xt){\partial _\mu }{G_\rho }(xt,x't')} \right.  \cr
&{}& \left. { - \int {dx'} \int\limits_0^t {dt'} {\Sigma ^\rho }(x't',xt){\partial _\mu }{G_K}(xt,x't')} \right)\label{b4}
\end{eqnarray}
Note that $t_0=0$ as the cosine potential and hence the self-energies are non-zero only after this time.
The second term of Eq.~(\ref{divergenceTmunu}) is
\begin{eqnarray}
&{}&  {1 \over {8i}}\int {dx'} dt'\left[ { - {\Sigma ^K}(xt,x't'){\partial _\mu }{G_\rho }(x't',xt){\rm{sg}}{{\rm{n}}_C}(t' - t)} \right.  \cr
  &{}& \left. { - {\Sigma ^\rho }(xt,x't'){\partial _\mu }{G_K}(x't',xt){\rm{sg}}{{\rm{n}}_{\rm{C}}}(t - t')} \right]  \cr
  &{}&  =  {1 \over {8i}}\int {dx'} dt'\left[ {{\Sigma ^K}(x't',xt){\partial _\mu }{G_\rho }(xt,x't'){\rm{sg}}{{\rm{n}}_C}(t' - t)} \right.  \cr
  &{}& \left. { + {\Sigma ^\rho }(x't',xt){\partial _\mu }{G_K}(xt,x't'){\rm{sg}}{{\rm{n}}_{\rm{C}}}(t - t')} \right]  \cr
  &{}&  = -{1 \over {4i}}\left( {\int {dx'} \int\limits_0^t {dt'} {\Sigma ^K}(x't',xt){\partial _\mu }{G_\rho }(xt,x't')} \right.  \cr
  &{}& \left. { - \int {dx'} \int\limits_0^t {dt'} {\Sigma ^\rho }(x't',xt){\partial _\mu }{G_K}(xt,x't')} \right) \label{b5}.
\end{eqnarray}
The terms in Eqs.~(\ref{b4}) and~(\ref{b5}) are equal, and summing them we obtain Eq.~(\ref{dTmunuKandrhocomp}).

\section{Proof of Eq.~(\ref{zm})} \label{zeromode}

Here we will show that $\frac{dT^{00}_{\Sigma}(T)}{dT}=0$. Since from
energy conservation $\frac{dE}{dT}=0$, this
implies that $\frac{dT^{00}_{\rm kin}(T)}{dT}=0$.
Using Eq.~(\ref{tsig}), we find that
\begin{eqnarray}
&&\frac{dT_{\Sigma}^{00}}{dT}= -\frac{uK}{4}\int_0^{\infty}dqe^{-\alpha q}\left[i\Sigma^K(q,T)\right.\nonumber \\
&&\left. -i\left(\Sigma^R-\Sigma^A\right)(q,T)F(q,T)\right]
\end{eqnarray}
Using Eq.~(\ref{SigK}),
\begin{eqnarray}
&&\int_0^{\infty}dq e^{-\alpha q}i\Sigma^K = {\left( {{{gu\gamma } \over {{\alpha ^2}}}} \right)^2}\int\limits_{ - \infty }^\infty  {d\tau }
\int\limits_{ - \infty }^\infty  {dr}   \nonumber \\
&&\times {e^{ - {I}(T,r,\tau )}}\cos \left[K_{\rm eq}\sum\limits_{\varepsilon  =  \pm 1} {{{\tan }^{ - 1}}
\left( {{{u\tau  + \varepsilon r} \over \alpha }} \right)} \right] \nonumber \\
&&\times \int_0^{\infty}dqe^{-\alpha q} \cos (qr)\cos (u q \tau )
\end{eqnarray}
Using $\int_0^{\infty}dqe^{-\alpha q} \cos (qr)\cos (u q \tau ) = \frac{1}{2\alpha}
\sum\limits_{\epsilon=\pm}\frac{1}{1+(u\tau+\epsilon r)^2/\alpha^2}$ and that
\begin{eqnarray}
&&\frac{\alpha}{K_{\rm eq}}\frac{\partial }{\partial (u\tau +\epsilon r)}\sin\left[K_{\rm eq}
\sum\limits_{\epsilon'=\pm}\tan^{-1}\left(\frac{u\tau+\epsilon' r}{\alpha}\right)\right]=\nonumber \\
&&\cos\left[K_{\rm eq}
\sum\limits_{\epsilon'=\pm}\tan^{-1}\left(\frac{u\tau+\epsilon' r}{\alpha}\right)\right]
\left[\frac{1}{1+(u\tau +\epsilon r)^2/\alpha^2}\right]\nonumber
\end{eqnarray}
we find,
\begin{eqnarray}
&&\int_0^{\infty}dq e^{-\alpha q}i\Sigma^K = {\left( {{{gu\gamma } \over {{\alpha ^2}}}} \right)^2}\int\limits_{ - \infty }^\infty  {d\tau }
\int\limits_{ - \infty }^\infty  {dr} {e^{ - {I}(T,r,\tau )}}\nonumber \\
&&\frac{1}{2K_{\rm eq}}\sum\limits_{\epsilon'=\pm}\frac{\partial }{\partial (u\tau +\epsilon' r)}
\sin \left[K_{\rm eq}\sum\limits_{\varepsilon  =  \pm } {{{\tan }^{ - 1}}
\left( {{{u\tau  + \varepsilon r} \over \alpha }} \right)} \right]\nonumber\\
\end{eqnarray}
Now one may integrate the above expression by parts and use that
$
\frac{\partial }{\partial (u\tau +\epsilon r)}I(r,\tau,T) = K_{\rm eq}\int_0^{\infty}dq e^{-\alpha q}F(q,T)
\sin\left[q(u\tau + \epsilon r)\right]$.
This leads to
\begin{eqnarray}
&&\int_0^{\infty}dq e^{-\alpha q}i\Sigma^K = {\left( {{{gu\gamma } \over {{\alpha ^2}}}} \right)^2}\int\limits_{ - \infty }^\infty  {d\tau }
\int\limits_{ - \infty }^\infty  {dr}   \nonumber \\
&&\times {e^{ - {I}(T,r,\tau )}}\sin \left[K_{\rm eq}\sum\limits_{\varepsilon  =  \pm } {{{\tan }^{ - 1}}
\left( {{{u\tau  + \varepsilon r} \over \alpha }} \right)} \right] \nonumber \\
&&\times \int_0^{\infty}dqe^{-\alpha q} \cos (qr)\sin (u q \tau )F(q,T)\nonumber \\
&&= \int_0^{\infty}dq e^{-\alpha q}i\left[\Sigma^R-\Sigma^A\right](q,T)F(q,T)
\end{eqnarray}
Thus we have proved Eq.~(\ref{zm}).

\section{Perturbative evaluation of the inelastic scattering rate} \label{inelasticrate}

In this section we discuss the inelastic scattering rate or dissipation defined in Eq.~(\ref{gamdef}) for two cases, one is for
small deviations of the system from thermal equilibrium at a temperature $T_{\rm eq}$,~\cite{Sirker11}
and the second is by doing perturbation theory for the quantum quench.~\cite{Mitra11,Mitra12a}

In equilibrium and finite temperature, $K_{\rm eq}=K_{\rm neq}$ and
and the distribution function of the bosons is $\langle 2b_{p}^{\dagger}b_p + 1\rangle=\coth\frac{u|p|}{2T_{\rm eq}}$. This leads to
the following inelastic scattering rate at temperature $T_{\rm eq}$, (in units of the cut-off)
\begin{eqnarray}
&&\eta_{T} =\left(\frac{\pi K}{2}\right)g^2 \gamma^2
\int_{-\infty}^{\infty}dr \int_{-\infty}^{\infty}d\tau \nonumber \\
&&\tau \sin\left[K_{\rm eq}\tan^{-1}(\tau+r)
 +K_{\rm eq}\tan^{-1}(\tau-r)\right]\nonumber \\
&&e^{-K_{\rm eq}\left[f_T(\tau+r)+f_T(\tau-r)\right]}\label{gamT1}
\end{eqnarray}
where
\begin{eqnarray}
&&f_T(x) = \int_0^{\infty}dq e^{-q}\left[1-\cos(qx)\right]\coth\left(\frac{uq}{2T_{\rm eq}\alpha}\right)\nonumber\\
&&=\ln(\sqrt{1+x^2})+2\ln\Gamma\left(1+\frac{\alpha T_{\rm eq}}{u}\right)\nonumber \\
&&-\ln\Gamma\left(1+\frac{\alpha T_{\rm eq}}{u}-i\frac{\alpha T_{\rm eq}x}{u}\right)\nonumber \\
&& -\ln\Gamma\left(1+\frac{\alpha T_{\rm eq}}{u}+i\frac{\alpha T_{\rm eq}x}{u}\right)
\end{eqnarray}
and $\Gamma$ is the Gamma function.
For $x\gg1$
\begin{eqnarray}
f_T(x\gg 1) = \ln\left[\frac{u}{\alpha\pi T_{\rm eq}}\sinh\left(\frac{\pi \alpha T_{\rm eq} x}{u}\right)\right]
\end{eqnarray}
From this it follows that the dissipation at finite temperature scales with temperature as follows,
\begin{eqnarray}
\eta_T \propto g^2\gamma^2
\left(\alpha T_{\rm eq}/u\right)^{2K_{\rm eq}-3}
\end{eqnarray}
in agreement with Ref.~\onlinecite{Sirker11} if we set $\gamma = 2\sqrt{2}$.

Let us now consider the case of the quantum quench. Using bare correlators ($g=0$) for the expectation values
$\langle \phi \phi \rangle$ (see Section~\ref{quench}),
explicit expressions for the self-energy in Eqns.~(\ref{piR}) and~(\ref{piK}) are given below,
\begin{eqnarray}
&&\Pi^R(x_1t_1,x_2t_2)= -\left(\frac{gu\gamma}{\alpha^2}\right)^2\theta(t_1-t_2)\nonumber \\
&&\sin\left[K_{\rm eq}\tan^{-1}\left(\frac{u(t_1-t_2)+(x_1-x_2)}{\alpha}\right)\right. \nonumber \\
&&\left. + K_{\rm eq}
\tan^{-1}\left(\frac{u(t_1-t_2)-(x_1-x_2)}{\alpha}\right)
\right]\nonumber \\
&&\times f_{\rm ss}(x_1-x_2,t_1-t_2) f_{\rm tr}(x_1-x_2,t_1,t_2)
\label{PiR}
\end{eqnarray}
and
\begin{eqnarray}
&&\Pi^K(x_1t_1,x_2t_2)=-i\left(\frac{gu\gamma}{\alpha^2}\right)^2\nonumber \\
&&\cos\left[K_{\rm eq}\tan^{-1}\left(\frac{u(t_1-t_2)+(x_1-x_2)}{\alpha}\right)\right. \nonumber \\
&&\left. + K_{\rm eq}
\tan^{-1}\left(\frac{u(t_1-t_2)-(x_1-x_2)}{\alpha}\right)
\right]\nonumber \\
&&\times f_{\rm ss}(x_1-x_2,t_1-t_2) f_{\rm tr}(x_1-x_2,t_1,t_2)
\label{PiK}
\end{eqnarray}
where $f_{\rm ss}$ is a function that is translationally invariant in time, while $f_{\rm tr}$ contains transients.
Their explicit expressions are,
\begin{eqnarray}
&&f_{\rm ss}(x_1-x_2,t_1-t_2)=
\left[\sqrt{\frac{\alpha^2}{\alpha^2 + \left[u(t_1-t_2)+(x_1-x_2)\right]^2}}\right.\nonumber\\
&&\left. \times \sqrt{
\frac{\alpha^2}{\alpha^2 + \left[u(t_1-t_2)-(x_1-x_2)\right]^2}}\right]^{K_{\rm neq}}\label{fss}\\
&&f_{\rm tr}(x_1-x_2,t_1,t_2)=
\left[\sqrt{\frac{\alpha^2 +(2ut_1)^2}{\alpha^2 + \left[u(t_1+t_2)+(x_1-x_2)\right]^2}}\right.\nonumber\\
&&\left. \times \sqrt{
\frac{\alpha^2+ (2ut_2)^2}{\alpha^2 + \left[u(t_1+t_2)-(x_1-x_2)\right]^2}}\right]^{K_{tr}}
\label{ftr}
\end{eqnarray}
Above $K_{\rm eq} = \frac{\gamma^2}{4}K$,
$K_{\rm neq} = \frac{\gamma^2}{8}K_0\left(1+ \frac{K^2}{K_0^2}\right)$, and
$K_{\rm tr}= \frac{\gamma^2}{8}K_0\left(1- \frac{K^2}{K_0^2}\right)$.
At very long times after the quench, we may set $(t_1+ t_2)/2\rightarrow \infty$ while keeping $t_1-t_2$ arbitrary.
In this case $f_{tr}\rightarrow 1$ and the
the scattering rates may be evaluated analytically to give,
\begin{eqnarray}
&&\eta_0 = \left(\frac{\pi K}{2}\right)g^2\gamma^2\nonumber\\
&&\times \left[\frac{\pi}{2^{K_{\rm neq}-1}(K_{\rm neq}-1)
B\left(\frac{K_{\rm eq}+K_{\rm neq}}{2},\frac{K_{\rm neq}-K_{\rm eq}}{2}\right)}\right]\nonumber \\
&&\times \left[\frac{\pi}{2^{K_{\rm neq}-2}(K_{\rm neq}-2)
B\left(\frac{K_{\rm eq}+K_{\rm neq}-2}{2},\frac{K_{\rm neq}-K_{\rm eq}}{2}\right)}  \right. \nonumber \\
&&\left. - \frac{\pi}{2^{K_{\rm neq}-2}(K_{\rm neq}-2)
B\left(\frac{K_{\rm eq}+K_{\rm neq}}{2},\frac{K_{\rm neq}-K_{\rm eq}-2}{2}\right)}\right]\nonumber \\
\label{IetaA}
\end{eqnarray}

Another useful quantity is the strength of the noise which is given by the correlation function~\cite{Mitra11,Mitra12a} $\Pi^K$.
Its expression at long-wavelengths
is given by,
\begin{eqnarray}
&&I_{K} = i\int_{-\infty}^{\infty}d\tau \int_{-\infty}^{\infty} dr \nonumber\\
&&\times \Pi_K(r,\tau=t_1-t_2,\frac{t_1+t_2}{2}\rightarrow \infty)
\end{eqnarray}
Within perturbation theory we obtain,
\begin{eqnarray}
&&I_K= 2g^2\gamma^2\times \nonumber \\
&&\left[\frac{\pi}{2^{K_{\rm neq}-1}(K_{\rm neq}-1)
B\left(\frac{K_{\rm eq}+K_{\rm neq}}{2},\frac{K_{\rm neq}-K_{\rm eq}}{2}\right)}\right]^2\label{IT0}
\end{eqnarray}

The ratio of the noise in Eq.~(\ref{IT0}) and the dissipation in Eq.~(\ref{IetaA}) may be used to define an effective
temperature,
\begin{eqnarray}
{\tilde{T}_{{\rm{eff}}}} = {u \over \alpha }\left( {{{{K_{{\rm{neq}}}} - 2} \over {2{K_{{\rm{eq}}}}}}} \right)
\end{eqnarray}
This effective-temperature does not have any physical meaning as the system is out of equilibrium,
and different combinations of response and correlation functions in general will give different effective-temperatures.
However, it is interesting to note that in the limit of large quenches $K_{\rm neq}\gg K_{\rm eq}$, this temperature approaches
the true temperature $T_{\rm eq}$ (Eq.~(\ref{TempLarge})) the system would reach if it thermalized.

\end{document}